%% file: gspice.tex
\documentclass[trackchanges]{aastex701}
\usepackage[utf8]{inputenc}
\usepackage{natbib}
\usepackage{graphicx}
\usepackage{dcolumn}
\usepackage{bm}
\usepackage{amsmath}
\usepackage{amstext}
\usepackage{CJK}
\usepackage{color}
\usepackage{hyperref}
\usepackage[ruled,vlined]{algorithm2e}
\usepackage{todonotes} % for comments

\newcommand{\doit}[1]{\todo[inline,color=green!20!white]{\textbf{To-do:} #1}}
\usepackage{soul}
\input{bangla_commands}
\def\deg{\ifmmode^\circ\else$^\circ$\fi}
\def\arcsec{\ifmmode^{\prime\prime}\else$^{\prime\prime}$\fi}
\def\arcmin{\ifmmode^{\prime}\else$^{\prime}$\fi}
\def\GSPICE{\texttt{GSPICE}}

\begin{document}

\title{Data-Driven Stellar Spectral Modelling with {\GSPICE}}

%\author{Douglas P. Finkbeiner\altaffilmark{1,2,3}}
\author[0000-0003-2808-275X]{Douglas P. Finkbeiner}
\affiliation{Center for Astrophysics\:\textbar\:Harvard \& Smithsonian, 
60 Garden St., Cambridge, MA 02138, USA}
\affiliation{Department of Physics, 
Harvard University, 17 Oxford St, Cambridge, MA 02138}
\email{dfinkbeiner@cfa.harvard.edu}
%\affiliation{Max-Planck-Institut f\"ur Astronomie, K\"onigstuhl 17, D-69117 Heidelberg, Germany}

\author[0000-0003-2573-9832]{Joshua S. Speagle (\begin{CJK*}{UTF8}{gbsn}沈佳士\ignorespacesafterend\end{CJK*})}
\affiliation{Department of Statistical Sciences, University of Toronto, 9th Floor, Ontario Power Building, 700 University Ave, Toronto, ON M5G 1Z5, Canada}
\affiliation{David A. Dunlap Department of Astronomy \& Astrophysics, University of Toronto, 50 St George Street, Toronto, ON M5S 3H4, Canada}
\affiliation{Dunlap Institute for Astronomy \& Astrophysics, University of Toronto, 50 St George Street, Toronto, ON M5S 3H4, Canada}
\affiliation{Data Sciences Institute, University of Toronto, 17th Floor, Ontario Power Building, 700 University Ave, Toronto, ON M5G 1Z5, Canada}
\email[show]{j.speagle@utoronto.ca}

\author[0000-0002-5652-8870]{Tanveer Karim ({\bng tanviir kirm})}
\affiliation{David A. Dunlap Department of Astronomy \& Astrophysics, University of Toronto, 50 St George Street, Toronto, ON M5S 3H4, Canada}
\affiliation{Dunlap Institute for Astronomy \& Astrophysics, University of Toronto, 50 St George Street, Toronto, ON M5S 3H4, Canada}
\affiliation{Center for Astronomy, Space Science and Astrophysics, Independent University, Bangladesh, Dhaka 1229, Bangladesh}
\email{tanveer.karim@utoronto.ca}
% \affiliation{Center for Astrophysics\:\textbar\:Harvard \& Smithsonian, 
% 60 Garden St., Cambridge, MA, USA 02138}

% \altaffiltext{1}{Institute for Theory and Computation,
%   Harvard-Smithsonian Center for Astrophysics, 
%   60 Garden Street, MS-51, Cambridge, MA 02138 USA} 

% \altaffiltext{2}{Department of Physics, 
%   Harvard University, Cambridge, MA 02138 USA}

% \altaffiltext{3}{Max-Planck-Institut f\"ur Astronomie, K\"onigstuhl 17,
%   D-69117 Heidelberg, Germany}

\begin{abstract}

Spectral data reduction pipelines deal with a wide variety of challenges including masking cosmic rays, calibrating wavelength solutions, and estimating background noise while trying to remain model-agnostic. Traditional methods rely on hardware-specific code or pre-calculated stellar model templates to solve this problem, making them model-dependent and not suitable for large datasets that may contain new classes of objects. To solve this problem, we present a flexible, data-driven method: the GausSian PIxelwise Conditional Estimator ({\GSPICE}) that models an ensemble of spectra as a multivariate Gaussian and estimates the expected value and expected variance of each pixel in each spectrum conditional on others. {\GSPICE} compares observed fluxes and errors to its own flux and error estimates to reveal outliers, which then can be completely masked or replaced by their estimates. We apply {\GSPICE} to 3.9 million stellar spectra from the LAMOST survey, and show that variations of the method can directly identify and correct both individual pixel-level outliers (e.g., from cosmic ray hits) as well as extended systematic features (e.g., from incorrect wavelength calibrations), while still providing a novel characterization of the true per-pixel measurement uncertainties. We also demonstrate how {\GSPICE} can take advantage of data partitioning with an application to diffuse interstellar bands. Implementations of {\GSPICE} in both Python and IDL can be found \href{here}{http://github.com/dfink/gspice}.\\
\end{abstract}

% GSPICE detects and repairs artifacts in spectral data caused by sensor defects and cosmic-ray hits, and is useful for validation of spectral hardware and data processing systems.  Unlike traditional methods that require hardware-specific code, GSPICE takes a data-driven approach, modeling an ensemble of spectra as a multivariate Gaussian and estimating the expected value of each pixel in each spectrum conditional on others.  Significant deviation of observed values from these estimates reveals outliers, which can be replaced by their estimates.  We provide an implementation of GSPICE in both Python and IDL, and show results based on 3.9 million stellar spectra from the LAMOST survey. 

%\tableofcontents
\section{Introduction}
\label{sec:intro}

Spectral data reduction is a fundamental challenge in modern astronomy. No spectroscopic dataset is perfect; observations are frequently plagued by sensor defects like bad CCD columns, transient artifacts such as cosmic-ray hits, or astrophysical interlopers like background galaxies. In fiber spectrographs, software failures can occasionally assign incorrect wavelength solutions to entire sections of a trace. Similarly, hyperspectral imaging (a 3-D dataset with a spectrum of every point in a 2-D image) suffers from complex optical distortions such as \textit{smile} and \textit{keystone} \citep{leung22}.

These challenges are becoming acute in the era of massive spectroscopic surveys such as DESI \citep{desi-main-cite-2022}, \textit{Gaia} \citep{gaia18}, SDSS V \citep{sdssv17}, and \textit{Euclid} \citep{euclid11}. With millions of spectra to process, manual inspection is impossible. Traditional reduction pipelines typically rely on template matching (comparing observations to theoretical models) to identify defects. However, these methods are brittle: they only find the errors they were designed to look for (the ``known unknowns''). Furthermore, when theoretical models are imperfect (as is often the case for high-redshift galaxies or complex stellar atmospheres), a valid scientific discovery might be flagged as a data error; conversely, a subtle instrument glitch might be accepted as real physics.

To address this, we need a flexible way to determine if an observation conforms to expectations without relying on rigid physical models. The intuition behind this approach is straightforward and builds on the fundamental fact that stellar spectra are highly redundant, with the flux at one wavelength strongly correlated with the flux at many others due to the underlying physics of atomic transitions. This means that if we know the shape of the spectrum in some subset of wavelengths, we should be able to predict the behavior at others with high precision. If the actual data deviates significantly from this prediction, we have likely found either a processing artifact or a potential outlier.

In addition to simply making a prediction, we also need to consider the relative variability of various spectral features to ensure that rare objects (which may not be represented in training data) still must be handled properly and not rejected outright. For instance, emission and absorption features from the atmosphere (e.g., sky lines) or other material along the line of sight (e.g., dust) may dominate the variance at some wavelengths, while other features might be intrinsically variable (e.g., metal-dependent absorption features). We therefore need an approach that is able to detect artifacts and repair them with accurate estimates of the mean \textit{and variance} for each potential outlier pixel. Correct variance estimates are also often important for downstream analysis.

In order to address both of these issues, in this work we introduce the GausSian PIxelwise Conditional Estimator ({\GSPICE}). {\GSPICE} uses the highly covariate nature of spectra to estimate the expected flux of an object at a pixel conditioned on the fluxes of other pixels. Similar to \citet{2022ApJ...933..155S}, the Gaussian assumption of the stellar and galactic spectra enables us to use exact and closed-form solutions for pixelwise flux estimation. Furthermore, our probabilistic approach shares a theoretical lineage with recent advancements in component separation, most notably MADGICS (Marginalized Analytic Data-generated Gaussian Inference for Component Separation; \cite{madgics2023}). Both GSPICE and MADGICS leverage the power of Gaussian priors to model astronomical signals. However, where MADGICS focuses on the marginalization of components to separate additive signals (e.g., separating dust emission from CMB, or tellurics from stellar spectra), GSPICE focuses on *conditional* inference at the pixel level.

In contrast to approaches like principal component analysis (PCA) that compress the spectrum into a linear combination of only a modest number of ``principal components'' (i.e., a small number of the eigenvectors of the covariance), {\GSPICE} is able to leverage the full covariance structure of the dataset. This not only makes the method more robust to outliers when training, but also allows for analysis of a spectrum on a pixel-by-pixel basis. By comparing the observed flux to this data-driven prediction, we can identify outliers with high sensitivity, whether they are sharp cosmic rays or broad calibration errors. Once identified, these outliers can be mathematically ``in-painted'' with their predicted values, effectively repairing the spectrum for downstream analysis.

% With this intuition, we introduce the GausSian PIxelwise Conditional Estimator (GSPICE) algorithm in this paper. GSPICE uses the highly covariate nature of spectra to estimate the expected flux of an object at a pixel conditioned on the fluxes of other pixels. The Gaussian assumption of the stellar and galactic spectra enables us to use exact and closed-form solutions for pixelwise flux estimation. Traditional approaches like PCA model each spectrum as a linear combination of only a modest number of principal components, i.e., the eigenvectors of the covariance, and determine these coefficients to reconstruct the spectrum. Moreover, the expected relative weighting of the principal components (related to the eigenvalues of the covariance) is ignored, and PCA approaches are sensitive to outliers. This is because PCA can be thought of as estimating bases where the $L_2$ norm is minimized; since outliers tend to dominate $L_2$ norms, PCAs by proxy are susceptible to their presence. In contrast, GSPICE does not transform coordinates of the data based on the linear combination of a subset of eigenvector, but rather uses all the eigenvectors exactly to provide the best reconstruction possible. Thus, leveraging the full covariance matrix makes GSPICE less susceptible to outliers. 

There are necessary limitations to this approach. To learn the correlations effectively, the spectra must be shifted to a common reference velocity (e.g., rest frame). Additionally, features that move relative to the star, such as foreground interstellar absorption, require special handling (although see Section \ref{sec:DIB}). Finally, the instrumental line-spread function at each wavelength must be approximately uniform across the sample.

In Section \ref{sec:method}, we detail the probabilistic framework of the algorithm and its relationship to Gaussian Processes (GPs) and other approaches in the literature. We apply GSPICE to 3.9 million spectra from the LAMOST survey \citep{LAMOST:2015} in Section \ref{sec:lamost} to demonstrate artifact detection and repair. In Section \ref{sec:DIB}, we demonstrate how the method can be used to isolate foreground signals. Section \ref{sec:concl} summarizes our results and future applications. 

Appendix \ref{sec:comp} describes the mathematical and computational methodology that allows us to compute a $4000\times4000$ matrix inverse times a vector in $<1$ ms on a CPU core.

%%%%%%%%%%%%%%%%%%%%%%%%%%%%%%%%%%%%%%%%%%%%%%%%%%%%%%%%
%%%%%%%%%%%%%%%%%%%%  SECTION 2  %%%%%%%%%%%%%%%%%%%%%%%
%%%%%%%%%%%%%%%%%%%%%%%%%%%%%%%%%%%%%%%%%%%%%%%%%%%%%%%%

\section{GausSian PIxelwise Conditional Estimation ({\GSPICE})}
\label{sec:method}

We now describe our method in additional detail. Section \ref{subsec:preamble} gives our notation and setup. Section \ref{subsec:pca} discusses standard approaches with principal component analysis (PCA). Section \ref{subsec:gaussian} introduces the concept of a Gaussian prior on spectrum space, and contrasts our approach more explicitly with Principal Component Analysis (PCA). Section \ref{subsec:gp} relates our discretized wavelength grid to a non-stationary Gaussian process, providing justification for interpolation of the covariance. In Section \ref{subsec:gspice}, 
we present the basic formulas for Gaussian conditional estimation, and in Section \ref{subsec:outlier} describe how we use GSPICE to identify and replace outliers.

\subsection{Notation and Definitions}
\label{subsec:preamble}

In the following, we consider a stellar spectrum\footnote{While many of the general techniques discussed here could apply to galaxies as well (provided they are redshifted into the rest-frame and resampled appropriately), we consider this beyond the scope of this work.} with $N_\lambda$ wavelength bins to be a point in an $N_\lambda$-dimensional vector space. Past work \citep[e.g.,][]{SDSSSpectro:2012} has shown that astronomical spectra are not uniformly distributed in this space, but rather occupy a rather low-dimensional subspace, typically 10-30 dimensions. One way of parameterizing this low-dimension subspace is with theoretical models \citep{MARCSgrid:2008,Synspec:2011,Dotter:2016,Choi:2016}, which explicitly allow the user to vary the spectrum as a function of $k$ stellar parameters $\boldsymbol{\theta}=\{\theta_1,\dots,\theta_k\}$ such as effective temperature $T_{\rm eff}$, surface gravity $\log g$, and metal abundances $[X/H]$. We can then parameterize a theoretical spectrum as
\begin{equation}
    \mathbf{S}_\lambda^\star(\boldsymbol{\theta}) = \{S_\lambda(\lambda_1; \boldsymbol{\theta}), \dots, S_\lambda(\lambda_{N_\lambda}); \boldsymbol{\theta})\}
\end{equation}
where $S_\lambda^\star(\lambda_i; \boldsymbol{\theta})$ is the (potentially continuum-normalized) predicted spectral flux density at wavelength $\lambda_i$ for a given set of stellar parameters $\boldsymbol{\theta}$.

We will assume an observed spectrum is just a noisy version of an (unknown) underlying spectrum, i.e.
\begin{equation}
    \mathbf{S}_\lambda = \mathbf{S}_\lambda^\star(\boldsymbol{\theta}) + \boldsymbol{\epsilon}
\end{equation}
where the noise
\begin{equation}
    \boldsymbol{\epsilon} \sim N(\mathbf{0}, \mathbf{\Sigma}_\epsilon)
\end{equation}
is assumed to be drawn from a multivariate Normal (Gaussian) distribution with zero mean and some covariance $\mathbf{\Sigma}_\epsilon$ (which can capture correlations in the underlying pixels that could arise from data reduction, atmospheric effects, etc.) with probability density function
\begin{equation} \label{eq:gaussian}
G(\mathbf{x}; \boldsymbol{\mu}, \mathbf{\Sigma}) = \frac{1}{\sqrt{2\pi |\mathbf{\Sigma}|}} \exp\left(- (\mathbf{x}-\boldsymbol{\mu})^T \mathbf{\Sigma}^{-1} (\mathbf{x}-\boldsymbol{\mu})\right)\, ,
\end{equation}
where $|\mathbf{\Sigma}|$ is the determinant of $\mathbf{\Sigma}$. In subsequent discussion, we will sometimes suppress the notation for $\boldsymbol{\theta}$ in the observed spectrum (as we did above) for convenience and to emphasize that it is not necessary to know what $\boldsymbol{\theta}$ actually is (or $\boldsymbol{\epsilon}$) to analyze stellar spectra.

If we wanted to compute the mean (expectation value) at a particular wavelength $\lambda_i$ across a population distribution $P(\boldsymbol{\theta})$ of stellar parameters, this would be 
\begin{align}
    \mu_i \equiv \mathbb{E}[S_i] &= \int S_\lambda(\lambda_i) \, P(\boldsymbol{\theta}) \, P(\epsilon_i) \, {\rm d}\boldsymbol{\theta} \, {\rm d}\epsilon_i \\
    &= \int S_\lambda^\star(\lambda_i; \boldsymbol{\theta}) P(\boldsymbol{\theta}) \, {\rm d}\boldsymbol{\theta}
\end{align}
where we have defined $S_i \equiv S_\lambda(\lambda_i)$ as convenient shorthand. The expectation is insensitive to noise here because $\epsilon_i$ is additive and has zero mean. However, it does depend on the exact distribution of the population of stellar types in the sample. 

The covariance between two wavelengths $\lambda_i$ and $\lambda_j$, by contrast, depends both on the distribution of stellar types as well as the exact noise properties since the expectation
\begin{align}
    \Sigma_ij &\equiv \mathbb{E}[(S_i - \mu_i)(S_j - \mu_j)] \\ 
    &= \int (S_i - \mu_i)(S_j - \mu_j) \, P(\boldsymbol{\theta}) \, P(\epsilon_i, \epsilon_j) \, {\rm d}\boldsymbol{\theta} \, {\rm d}\epsilon_i \, {\rm d}\epsilon_j
\end{align}
now has contributions from cross-terms. This also should make sense: even just considering the variance $\Sigma_ii$, we should expect contributions from the noise level $\epsilon_i$ as well as the distribution of stellar parameters $P(\boldsymbol{\theta})$. This holds true if there are correlations in the noise realizations between pixels ($\Sigma_{\epsilon,ij} \neq 0$) as well.

For a finite sample of $N_{\rm spec}$ objects, we usually estimate the underlying mean and covariance using the sample mean and sample covariance, respectively.

%==================================================================
%=  FIGURE Cond2d   fig:cond2d
%==================================================================
\begin{figure*}[t]
\begin{center}
\includegraphics[width=\textwidth]{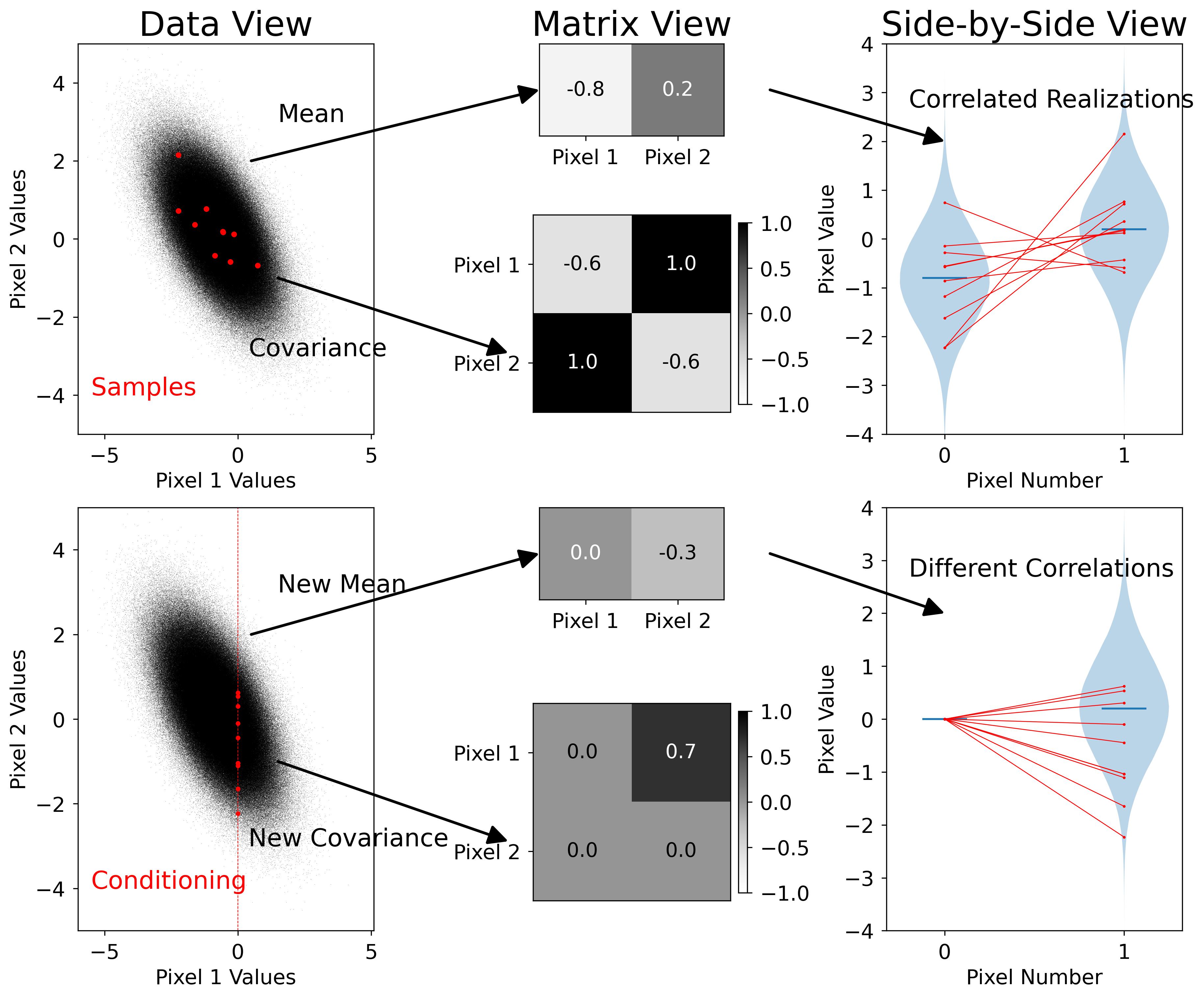}
\end{center}
\caption{Gaussian conditional estimate of $x_2$ given $x_1$.  A joint distribution $P(x_1,x_2)$ (gray ellipse, upper left panel) represents the probability distribution of many correlated realizations of $x_1$ and $x_2$ (red lines, upper right). The probability density of $x_2$ conditional on $x_1=0$ (red line, lower left panel), $P(x_2|x_1=0)$, is represented graphically (lower right) by the distribution of $x_2$ given the fixed value of $x_1=0$. See Section \ref{sec:method} for additional details.}
\label{fig:cond2d}
\end{figure*}

\subsection{Principal Component Analysis}
\label{subsec:pca}

If a large sample of spectra is available, a conventional data-driven approach is to apply principal component analysis \citep[PCA;][]{Pearson:1901,Hotelling:1933}. PCA relies on the fact that you can decompose the (mean-subtracted) sample covariance matrix $\mathbf{\Sigma}$ into a set of $N_{\lambda}$ eigenvectors $\{\mathbf{q}_1, \dots, \mathbf{q}_{N_\lambda}\}$ and eigenvalues $\{ \lambda_i \}$ as
\begin{equation}
    \mathbf{\Sigma} = \mathbf{Q}\boldsymbol{\Lambda}\mathbf{Q}^T = \sum_{i=1}^{N_\lambda} \lambda_i \mathbf{q_i} \mathbf{q_i}^T
\end{equation}
where $\mathbf{Q}$ is the stacked matrix of eigenvectors, $T$ is the transpose, and $\boldsymbol{\Lambda}$ is the diagonal matrix of eigenvalues. The individual terms $\lambda_{i} \mathbf{q}_{i} \mathbf{q}_{i}^{T}$ are rank-1 matrices representing the variance along the direction of the eigenvector $\mathbf{q}_{i}$. 

PCA selects the $N_{\rm PCA}$ eigenvectors with the largest eigenvalues as a basis for an $N_{\rm PCA}$-dimensional subspace within the larger spectral space. Assuming that we order the eigenvectors $\lambda_1 > \lambda_2 > \dots > \lambda_{N_\lambda}$, the PCA reconstruction of a spectrum is then simply its projection onto this truncated subspace
\begin{equation}
    \mathbf{S}_{\lambda, {\rm PCA}} = \sum_{i=1}^{N_{\rm PCA}} w_{i} \mathbf{q}_i = \left(\mathbf{q}_i^T \mathbf{S_\lambda}\right) \, \mathbf{q}_i
\end{equation}
The difference between the data and its projection defines a $\chi^2$-distributed test statistic
\begin{equation}
    \chi^2 = (\mathbf{S}_{\lambda, {\rm PCA}} - \mathbf{S}_\lambda)^T \mathbf{\Sigma}_{\epsilon} (\mathbf{S}_{\lambda, {\rm PCA}} - \mathbf{S}_\lambda)
\end{equation}
that can be evaluated (perhaps as a function of velocity) or inspected for outlier pixels in need of masking.

Although PCA is a simple and powerful tool, it has some noticeable shortcomings:
\begin{enumerate}
    \item Spectral features may be ignored if they are rare in the training data used to compute the covariance, and therefore do not appear in the first $N_{\rm PCA}$ eigenvectors.
    \item Criteria for the choice of $N_{\rm PCA}$ are vague.
    \item PCA discards the information about the relative importance of the eigenvectors, which may vary by several orders of magnitude.
\end{enumerate}
Because of this, the uncertainty of a PCA reconstruction is hard to assess. While it is possible to compute PCA reconstructions in the presence of noisy and/or missing data \citep{bailey2012pca}, interpreting the results on a per-spectrum basis can still be significantly challenging.

There are several additional drawbacks to PCA also worth noting. The first is that it is sensitive to outliers. This is because PCA can be thought of as estimating bases where the $L_2$ norm (sum of squares) is minimized. However, since outliers tend to dominate $L_2$ norms, they can have a large impact on eigenvalue ordering and eigenvector directions, leading PCAs by proxy to be susceptible to their presence.

The second is that, because it assumes that components are separable directly in wavelength-space, it cannot account for effects such as changes in velocity dispersions (convolutions) or shifts (translations). These instead manifest as ``ringing'' artifacts that spread out across many PCs that attempt to correct for these effects in a perturbative fashion \citep{2020AJ....160...45P}.

The last potential issue is that PCA does not naively give an independent estimator for the variance on a pixel-by-pixel basis.\footnote{By variance, we mean the measurement uncertainty itself. In other words, we are referring to the variance in the \textit{data}, not the variance in our estimate of the \textit{mean}. For the latter, we could in theory use a similar procedure as the one defined in Section \ref{subsec:gspice} when computing projections over the PCs.} By definition, noise is incompressible no matter what projection you choose; this means that it spreads out power among many of the low-eigenvalue components. While in the best case this means that PCA should ``de-noise'' the data, this removal of measurement noise information actually prevents us from investigating the properties of the measurement noise itself.

To try and address these concerns, we now turn our attention to approaches that can utilize the full-rank covariance matrix.

\subsection{Stellar Spectra as a Gaussian Distribution}
\label{subsec:gaussian}

One way we can use the entire covariance matrix is to assume that all appropriately normalized and mean-subtracted spectra are actually drawn from a Gaussian distribution whose covariance is identical to the one shown above
\begin{equation}
    \mathbf{S}_\lambda \sim N(\mathbf{0}, \mathbf{\Sigma})
\end{equation}
where again $\mathbf{\Sigma}$ is the $N_\lambda \times N_\lambda$ covariance matrix across all spectra in our sample (which has the same dimensions but is different from the specific measurement error $\mathbf{\Sigma}_\epsilon$ for a particular object; see Section \ref{subsec:preamble}).
In the most direct interpretation, this assumes that all spectra are distributed as an $N_\lambda$-dimensional Gaussian in spectral space. In practice, this does not have to be true in detail.\footnote{In fact, in practice this can be (and is) violated quite severely since stellar spectra samples tend to be multimodal. We will return to this point later.} Regardless of exact interpretation, however, this approach retains the full covariance structure of the training sample.

Assuming a joint Gaussian structure allows reconstruction of missing data. Splitting our data into two chunks and defining the associated means and covariances as
\begin{equation}
    \boldsymbol{\mu} = \left[
    \begin{matrix}
    \boldsymbol{\mu}_{1} \\ \boldsymbol{\mu}_{2}
    \end{matrix}
    \right],\quad 
    \mathbf{\Sigma} = \left[
    \begin{matrix}
    \mathbf{\Sigma}_{11} & \mathbf{\Sigma}_{12} \\ \mathbf{\Sigma}_{21} & \mathbf{\Sigma}_{22}
    \end{matrix}
    \right]
\end{equation}
we can then estimate the mean conditioned on $\mathbf{x}_2$ via
\begin{equation}
    \boldsymbol{\mu}_{1|2} = \mu_{1} + \mathbf{\Sigma}_{12} \mathbf{\Sigma}_{22}^{-1} (\mathbf{x}_{2} - \boldsymbol{\mu}_{2})
\end{equation}
\textit{and} the conditional covariance as
\begin{equation}
    \mathbf{\Sigma}_{1|2} = \mathbf{\Sigma}_{11} - \mathbf{\Sigma}_{12} \mathbf{\Sigma}_{22}^{-1} \mathbf{\Sigma}_{21}
\end{equation}
An example of what this procedure looks like across two pixels is shown in Figure \ref{fig:cond2d}.

Our usage of the entire eigenbasis (i.e. the full covariance matrix) allows us to retain all noise information, respect rare objects according to their prevalence in the training data, and also account for non-linear effects in the spectrum, such as velocity dispersion or shifts. This is again because the full suite of eigenvectors allows us to retain all noise/higher-order information for a given object (as opposed to PCA, where this information is lost). This is also helpful for outliers because the \textit{relative} projected coefficients of rare objects onto the eigenbasis can differ dramatically from the overall trends across the sample, and thus keeping lower-amplitude eigenvectors allows these objects to be properly represented by the data compared with a truncated basis (where their representations can be entirely lost).

%==================================================================
%=  FIGURE Covariance   fig:cov
%==================================================================
\begin{figure*}[t]
\begin{center}
\includegraphics[width=3.2in]{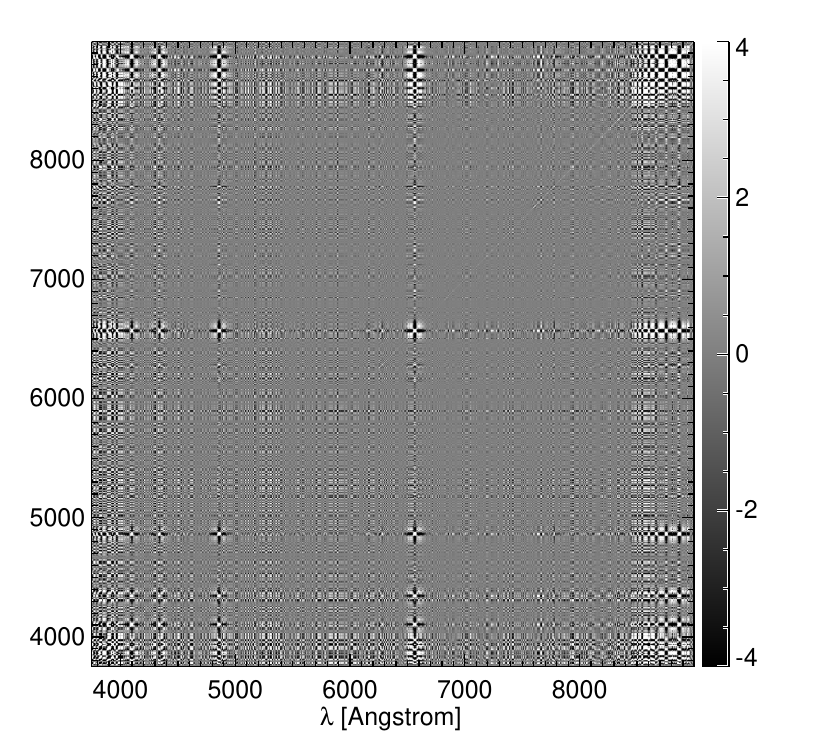}
\includegraphics[width=3.2in]{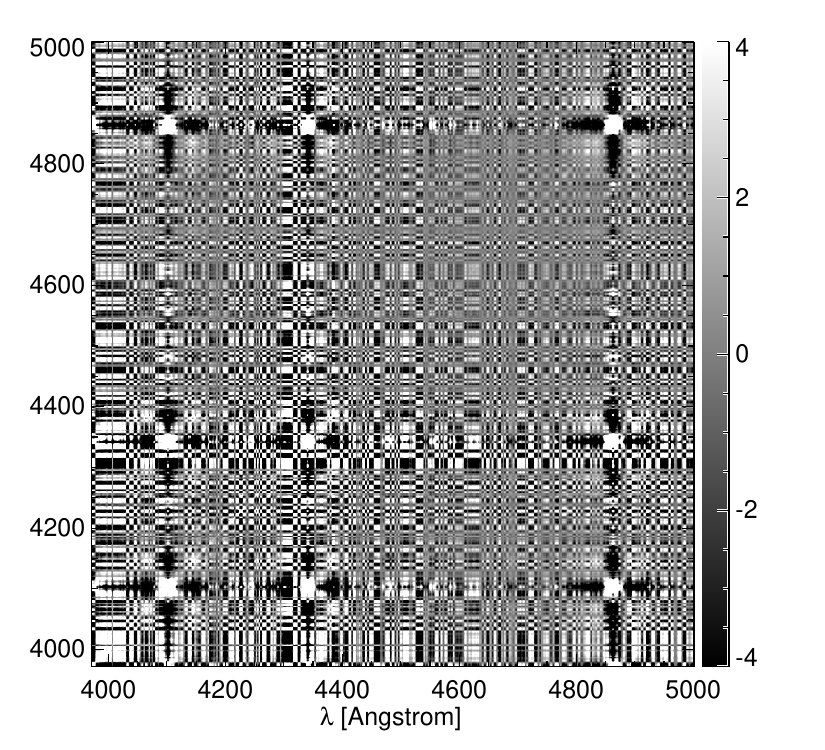}
\end{center}
\caption{Covariance matrix of stellar spectra for $\lambda=3800-9000$\AA\ (left) and a more limited wavelength range (right), as described in Section \ref{sec:lamost}. The most obvious bright spots in the right panel are H$\delta$ (4100), H$\gamma$ (4340), and H$\beta$ (4860). Bright spots have a dark cross-halo because of the continuum normalization.  
%\textbf{TODO: Move to LAMOST section??}
}
\label{fig:cov}
\end{figure*}

As mentioned above, one concrete benefit of this approach is that we can naturally incorporate measurement noise directly into the inference procedure. Naively, we could imagine doing so simply by using the combined covariance matrix $\mathbf{\Sigma} + \mathbf{\Sigma}_\epsilon$ when making predictions for the object in question, which would add in errors from the entire population covariance along with the individual object's measurement uncertainties. However, recall that, in general, this would require that we try to estimate the underlying population $\mathbf{\Sigma}$ by ``deconvolving'' the observed measurement uncertainties \citep{2011AnApS...5.1657B}, which assumes that the estimated measurement uncertainties are correct. Instead, we can actually probe the estimated measurement errors directly by using the \textit{actual variability} in the noisy measurements in an individual spectrum $\mathbf{S}_\lambda$ as well as across the entire sample in $\mathbf{\Sigma}$.

To make all of these points clearer, it can be helpful to recast conditional Gaussian estimation more directly in terms of the original eigenbasis. Taking our previous eigendecomposition $\mathbf{\Sigma} = \mathbf{Q} \mathbf{\Lambda} \mathbf{Q}^T$, we can do the same block-wise decomposition as above to get:
\begin{equation}
    \mathbf{Q} = \left[
    \begin{matrix}
    \mathbf{Q}_{11} & \mathbf{Q}_{12} \\ \mathbf{Q}_{21} & \mathbf{Q}_{22}
    \end{matrix}
    \right],\quad 
    \mathbf{\Lambda} = \left[
    \begin{matrix}
    \mathbf{\Lambda}_{1} & 0 \\ 
    0 & \mathbf{\Lambda}_{2}
    \end{matrix}
    \right]
\end{equation}
We can then express the sub-matrices of $\mathbf{\Sigma}$ using these partitions:
\begin{align}
    \mathbf{\Sigma}_{12} &= (\mathbf{Q} \mathbf{\Lambda}  \mathbf{Q}^{T})_{12} = \mathbf{Q}_{11} \mathbf{\Lambda}_{1} \mathbf{Q}_{21}^{T} + \mathbf{Q}_{12} \mathbf{\Lambda}_{2} \mathbf{Q}_{22}^{T}
    \\
    \mathbf{\Sigma}_{22} &= (\mathbf{Q} \mathbf{\Lambda}  \mathbf{Q}^{T})_{22} = \mathbf{Q}_{21} \mathbf{\Lambda}_{1} \mathbf{Q}_{21}^{T} + \mathbf{Q}_{22} \mathbf{\Lambda}_{2} \mathbf{Q}_{22}^{T}
\end{align}

This highlights that our conditional Gaussian prediction of the mean is essentially doing the following operation (from right to left):
\begin{enumerate}
    \item Calculate the deviation of the observed data from block 2 from the mean of block 2.
    \item Rotate the residual from block 2 into the global eigenspace (i.e. onto the principal components).
    \item Scale each projected component of block 2 by its relative importance to predicting the other components in block 1 and rotate the predicted deviations back into the observed data space.
    \item Apply as a deviation from the mean of block 1.
\end{enumerate}
We can also interpret the covariance estimation a similar way, where the covariance is reduced by the amount of mutual information shared between the two blocks projected through the global eigenbasis. The ability to use the entire eigenbasis for these projections and the relative rescaling based on the predicted quantities is what makes this approach more flexible and more powerful for spectral prediction than PCA-based methods.

%==================================================================
%=  FIGURE CondHetPop   fig:condhetpop
%==================================================================
\begin{figure*}[t]
\begin{center}
\includegraphics[width=0.84\textwidth]{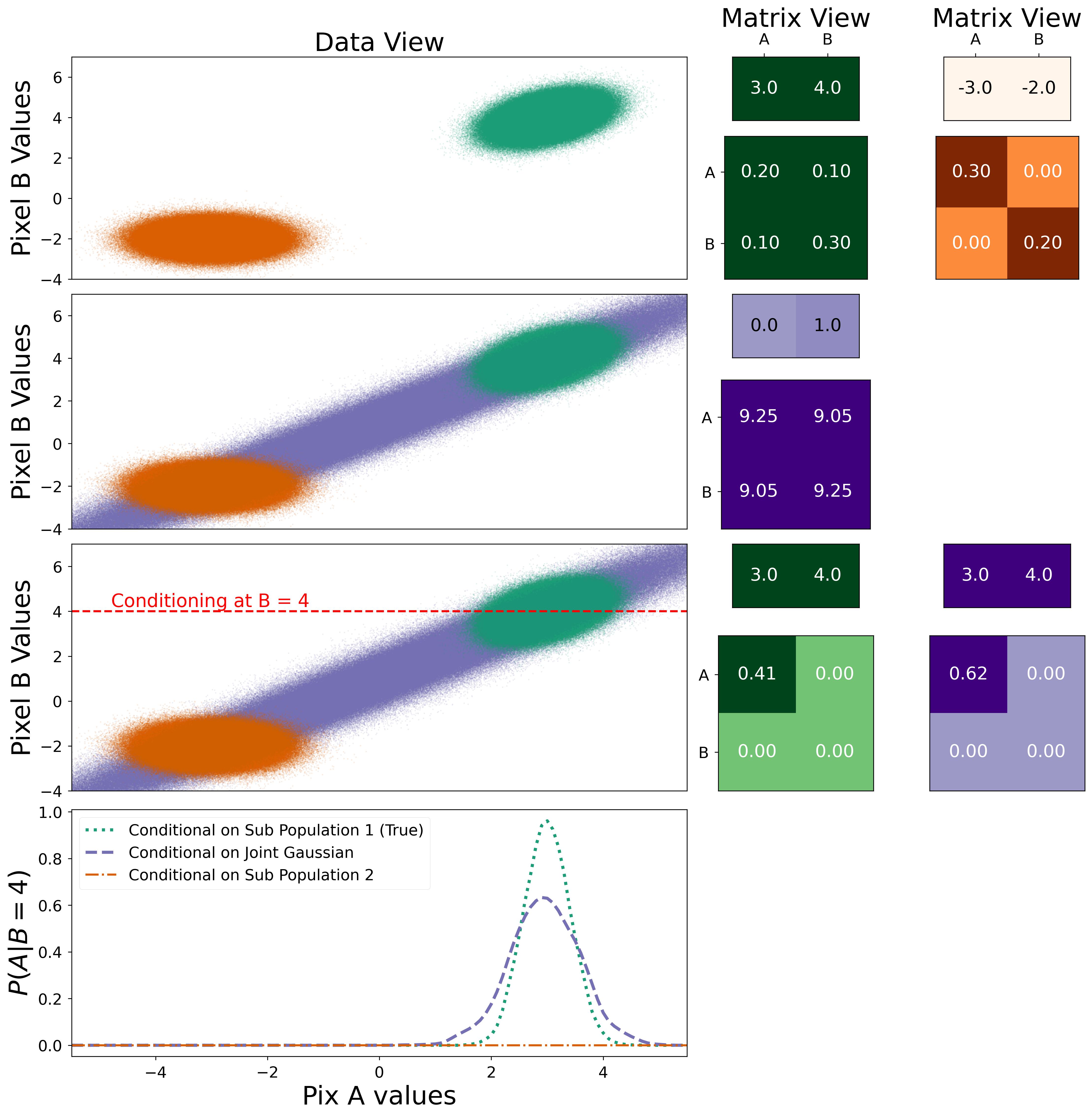}
\end{center}
\caption{An illustration showing how Gaussian conditional estimates can handle heterogeneous data. \textit{Top row}: Distribution of values of two spectral pixels (A and B) across a sample of two heterogeneous populations (green and orange). Their corresponding matrix view (shown in the same color) shows that their means (top two numbers) and covariances (bottom 2x2 numbers) are very different. \textit{Second row}: A single Gaussian model fit to the heterogeneous population (purple). \textit{Third row}: The estimates for the value of pixel A conditioning on pixel B (red dashed line) based on the true distribution and the joint Gaussian. \textit{Bottom row}: Conditional probability of pixel A given the value of pixel B (i.e. probability along the slice) for the two true populations and the joint Gaussian fit. Even if the joint Gaussian is overall a poor fit to the \textit{entire} population, its \textit{conditional} estimates nevertheless yield very reasonable answers.} %{\GSPICE} leverages this property to generate accurate predictions even while trained on a heterogeneous sample of stars.
\label{fig:condhetpop}
\end{figure*}

\subsection{Relation to a Continuous Gaussian Process}
\label{subsec:gp}

So far, we have considered samples of flux on a fixed wavelength grid, but the spectrum is fundamentally a continuous function of wavelength. The details of the sampling (wavelength and line-spread function, or LSF, at each grid point) may vary with each observation, but we must somehow interpolate to a uniform sampling at a reference velocity. How is it possible to interpolate a covariance matrix?

At a more abstract level, the covariance of interest is an infinite-dimensional quantity, a covariance of intensity at $\lambda_1$ and $\lambda_2$ for every ($\lambda_1,\lambda_2$) pair (though in practice we only have a discrete sampling of it). An infinite-dimensional Gaussian is known as a ``Gaussian process'' (GP; \citealt{RW:2006}). Similar to our original Gaussian distribution, these are now defined through a mean $\mu(\cdot)$ and covariance (kernel) function $K(\cdot, \cdot)$:
\begin{equation}
    S_\lambda(\lambda) \sim \mathcal{GP}\left(\mu(\lambda), K(\lambda, \lambda')\right)
\end{equation}
We can consider our finite-dimensional covariance matrix to be the discretized realization of such a latent GP. As long as we are considering the exact same set of wavelengths each time, working with a large covariance matrix is more or less identical to working with the latent Gaussian process.

Much has been done in recent years with \textit{stationary} GPs \citep{RW:2006}, meaning that the covariance can be specified by some function of the distance between data points rather than the exact positions of the data points, i.e.
\begin{equation}
    K(x,x')=K(|x-x'|)
\end{equation}
In the stationary case, the GP can be used to easily predict data in parts of parameter space -- in our case, at wavelengths -- that have not been observed. Stationarity is an adequate approximation for many phenomena, and recognition of the power of this approach, combined with increasingly capable computers and an influential textbook by \cite{RW:2006}, has greatly expanded the use of GPs in machine learning in recent years as well as in various fields of astrophysics.

Unfortunately, while a stationary GP is a powerful technique with wide applicability, it is not appropriate for spectra.  The dimensions of the spectral space correspond to particular wavelengths, and there are physical relationships between these wavelengths (not just the separation between them). For example, if two spectral bins separated widely in wavelength both contain an oxygen line, the values in those bins are correlated for a physical reason. In this case, we have a non-stationary GP with a rich covariance tying together disparate parts of the spectrum, and that structure is what allows us to estimate one spectral region conditional on another. Figure \ref{fig:cov} shows the rich structure present across real stellar spectra analyzed in Section \ref{sec:lamost}.

As long as the underlying GP is well sampled by our sampling grid, it is possible to make small adjustments to the sampling (e.g. Doppler shifts for a range of stellar velocities) by interpolation.\footnote{There is a danger that an interpolated covariance matrix will not be positive definite because of numerical noise. In practice, one builds a covariance matrix from interpolated data and applies it to interpolated data. It is safer to shift the data, not the matrix.} Once the underlying GP has been sampled onto our fixed set of wavelengths, we simply have an $N_\lambda \times N_\lambda$ covariance matrix that represents the discretized version of the underlying latent GP for spectra sampled at the $N_\lambda$ grid points (and it is no longer necessary to speak of GPs at all).

\subsection{Gaussian Conditional Estimation}
\label{subsec:gspice}

%\textbf{TODO: Clean up slightly. Add in gspice1 schematic figure and merge 2-D example directly into the text. Merge special cases directly into the text along with gspice2 and gspice3 figures to explain pixelwise and blockwise approach. Remove existing conditional estimation and pixelwise figures.}

The central goal of this work is to detect and repair bad or missing data in a spectral data set. Specifically, we estimate the values in some spectral pixels\footnote{We use ``pixel'' and ``wavelength bin'' interchangeably in this discussion.}, $k_*$, conditional on the values in other pixels, $k$.

% ========================================================
% ALGORITHM:     GSPICE_Covar
% ========================================================
\begin{algorithm}[t]
\textbf{Inputs:}
$X$ = scaled data, $N_\lambda \times N_{\rm s} $ \\
\KwResult{Sample covariance}
\vspace{0.1cm}
%\textbf{Notes:} 
\vspace{0.1cm}
\textit{Note: As an index, ``:" means ``all values."} \\

\vspace{0.2cm}
\SetAlgoLined
 \For{$i$ = 1 to $N_{\lambda}$}{
   $\mathbf{X}_{i,:} = \mathbf{X}_{i,:} - {\rm mean}(\mathbf{X}_{i,:}$)
 }
 $\mathbf{\Sigma} = \mathbf{X} \mathbf{X}^T / N_{\rm s}$ \;
 return $\mathbf{\Sigma}$
 \caption{\texttt{GSPICE\_covar}}
 \label{alg:gspicecovar}
\end{algorithm}

We will now represent each spectrum by a column vector, $\mathbf{x}$, and the value at indices $i$, $x_i$, with the $i$ index running over all wavelength bins.  The matrix $\mathbf{X}$ denotes a collection of spectra, and the $i^{\rm th}$ bin of the $s^{\rm th}$ spectrum is $X_{is}$.  The empirical sample covariance of the (mean-shifted) spectra $\mathbf{\Sigma} = \mathbf{X} \mathbf{X}^T / N_{\rm s}$ calculated according to Algorithm \ref{alg:gspicecovar}. In the notation of \cite{RW:2006}, the subset of pixels we condition on (``reference'' pixels), is again denoted $k$, and the pixels to be estimated, $k_*$. The covariance of reference pixels with themselves is $\mathbf{\Sigma}_{kk}$ and the covariance of the estimated pixels with the reference pixels $\Sigma_{k_*k}$.  Like any covariance matrix, $\mathbf{\Sigma}_{kk}$ is a symmetric positive semi-definite matrix, and its inverse can be calculated efficiently by Cholesky factorization. Most applications of the following equations do not even require an explicit inverse (see Appendix \ref{sec:comp} for computational details). 

The Gaussian conditional estimate (indicated with a tilde) of the values in pixels $k_*$ of a given spectrum is computed via (see also Section \ref{subsec:gaussian} for the general version with non-zero means):
\begin{equation}
\label{eq:condmean}
\tilde{\mathbf{x}}_{k_*} = \mathbf{\Sigma}_{k_*k}~ (\mathbf{\Sigma}_{kk})^{-1} \mathbf{x}_{k}\, .
\end{equation}
The matrix $\Sigma_{k_*k} \, (\Sigma_{kk})^{-1}$ may be computed once and applied to an entire array of spectra, yielding an array of estimates:
\begin{equation}
\label{eq:condmean2}
\tilde{\mathbf{X}}_{k_*i} = \mathbf{\Sigma}_{k_*k} \, (\mathbf{\Sigma}_{kk})^{-1} \mathbf{X}_{ki}\, ,
\end{equation}
where $\mathbf{X}_{ki}$ is the $N_k \times N_{\rm spec}$ array of spectra. The estimated variance again is 
\begin{equation}
\label{eq:condvar}
\tilde{\mathbf{\Sigma}}_{k_*k_*} = \mathbf{\Sigma}_{k_*k_*} - \mathbf{\Sigma}_{k_*k} \, (\mathbf{\Sigma}_{kk})^{-1} \mathbf{\Sigma}_{kk_*}.
\end{equation}
The estimated mean given by Eq.~(\ref{eq:condmean}) and calculated via Algorithm \ref{alg:gce} is again a linear function of $\mathbf{x}_k$, but the variance in Eq.~(\ref{eq:condvar}) again does not depend on $\mathbf{x}_k$. As we have alluded to obliquely in previous sections, this dependence on $\mathbf{x}_k$ means that some pre-scaling of the inputs will be necessary, both to have them contribute to the empirical covariance matrix with appropriate weight, and to recover realistic variance estimates.  In practice, we scale each spectrum by some measure of the $S/N$ ratio, estimate $\mathbf{\Sigma}$, and then scale the mean and variance estimates accordingly. In the LAMOST example in Section~\ref{sec:lamost}, we show an example of how the scaling is done following the procedure outlined in Section \ref{subsec:outlier}.

% ========================================================
% ALGORITHM:     GSPICE_gaussian_estimate
% ========================================================
\begin{algorithm}[t]
\textbf{Inputs:} \\
$k$ = index array of pixels to condition on \\
$k_*$ = index array of pixels to estimate \\ 
$\Sigma$ = scaled covariance matrix \\
$X$ = scaled data, $N_\lambda \times N_s $ \\
\KwResult{prediction and variance for pixels $k_*$ of every object.}
\vspace{0.1cm}

\textit{Note: As an index, ``:" means ``all values." The loop is done by matrix multiplication, and matrix inverses use the methodology in Appendix \ref{sec:comp}.}
\vspace{0.2cm}
\\
\SetAlgoLined
    $W \leftarrow \Sigma_{k_*k}~ (\Sigma_{kk})^{-1}$ \;
    $\tilde{\Sigma}_{k_*k_*} \leftarrow \Sigma_{k_*k_*} - \Sigma_{k_*k}~ (\Sigma_{kk})^{-1} \Sigma_{kk_*} $ \;
    \For{$s$ = 1 to $N_s$}{
        $\tilde{X}_{k_*,s} \leftarrow W X_{k,s}$ \;        
    }   
    return $\tilde{X}_{k_*,:}, \tilde{\Sigma}_{k_*k_*}$

 \caption{\texttt{GSPICE\_gaussian\_estimate}}
 \label{alg:gce}
\end{algorithm}

%==================================================================
%=  FIGURE Guard window   fig:nguard
%==================================================================
\begin{figure}[t!]
\begin{center}
\includegraphics[width=3.3in]{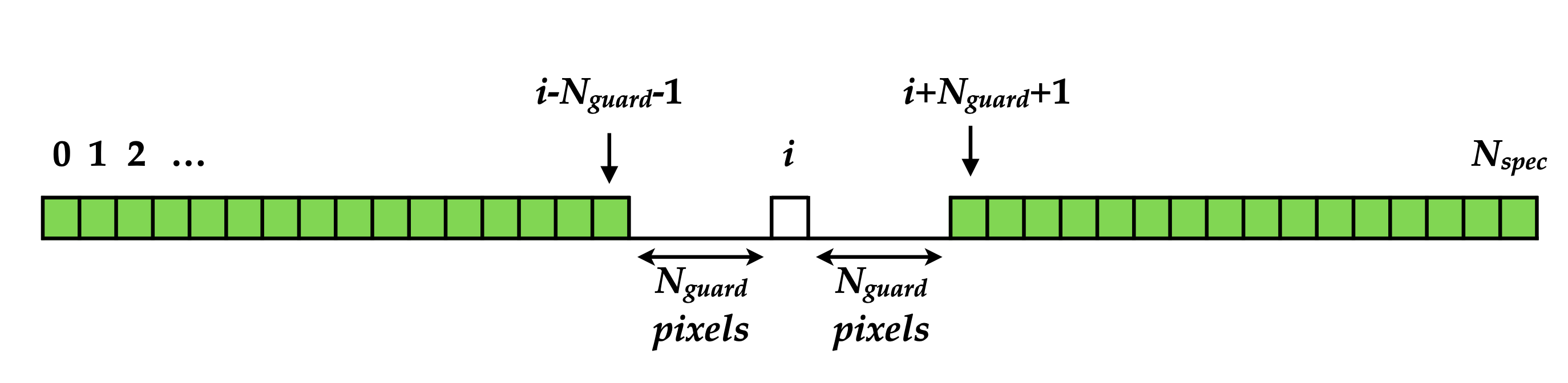}
\end{center}
\caption{Pixels used for Gaussian conditional estimation \textit{(green)}.  Pixel $i$ is predicted using all other pixels in the spectrum, except for a guard 
region around $i$ with length $N_{\rm guard}$.  In this work we take $N_{\rm guard}=20$, driven by the size of the smoothing kernel used for continuum normalization.}
\label{fig:nguard}
\end{figure}

We caution that although a covariance $\mathbf{\Sigma}_{kk}$ generated from noisy data is an unbiased estimate of the covariance, $\mathbf{\Sigma}_{kk}^{-1}$ is \textit{not} an unbiased estimator of the inverse covariance. The Hartlap correction \citep{Hartlap:2007} corrects for this bias\footnote{The expectation of the inverse covariance is larger than the simple inverse by a factor of $\langle\mathbf{\Sigma}^{-1}\rangle / \mathbf{\Sigma}^{-1} = (N_{\rm s}-1)/(N_{\rm s}-N_\lambda-2)$.} but in practice we usually have $N_{\rm s} \gg N_\lambda$ so the Hartlap correction is unimportant. In any case, this correction cancels in Eq. (\ref{eq:condmean}) and is only relevant for the variance estimate.

It may be difficult to visualize how Eqs.~(\ref{eq:condmean2}) and (\ref{eq:condvar}) play out in 4000 dimensions, especially given the fact that our underlying joint distribution of stellar spectra is almost certainly \textit{not} Gaussian. To build intuition, in Figure \ref{fig:condhetpop} we illustrate a Gaussian conditional estimate over a sample of two heterogeneous populations. In 2-D, assuming we have a bivariate Gaussian with some mean $\boldsymbol{\mu} = \{ \mu_1, \mu_2 \}$ and covariance $\mathbf{\Sigma}$, assuming
\begin{eqnarray}
\mu & = & 
\begin{bmatrix}
\mu_1 \\ \mu_2
\end{bmatrix} =
\begin{bmatrix}
0 \\ 0
\end{bmatrix} \\
\Sigma & = & 
\begin{bmatrix}
        \sigma_{11}^2 & \sigma_{12}^2 \\
        \sigma_{21}^2 & \sigma_{22}^2
\end{bmatrix}\, .
\end{eqnarray}
and substituting into Equations \ref{eq:condmean} and \ref{eq:condvar} we obtain a conditional distribution $P(x_2|x_1)= \mathcal{N}(\tilde{x}_2, \tilde{\Sigma}_{22})$ 
\begin{eqnarray}
\tilde{x}_2 & = & \sigma_{21}^2 \frac{1}{\sigma_{11}^2} x_1 \label{eq:mean2d}\\
\tilde{\sigma}^2_{22} & = & \sigma_{22}^2 - \sigma_{21}^2 \frac{1}{\sigma_{11}^2} \sigma_{12}^2\, . \label{eq:var2d}
\end{eqnarray}
This is simply the conditional probability, slicing the Gaussian along fixed $x_1$ to produce a lower-dimensional Gaussian with mean $\tilde{x}_2$ and variance $\tilde{\sigma}^2_{22}$.

\subsection{Outlier Identification}
\label{subsec:outlier}
The Gaussian conditional estimate provides a replacement value for bad pixels based on values in good pixels, but how do we know which pixels are bad? Simply evaluating $Z^2 = \mathbf{x}^T\mathbf{\Sigma}^{-1}\mathbf{x}$, for properly scaled $\mathbf{x}$ merely indicates that a spectrum as a whole is corrupted, but does not tell us what part is bad.\footnote{We distinguish between the test statistic $Z^2$ and the distribution $\chi^2_\nu$. $Z^2$ is distributed as $\chi^2_\nu$ for $\nu$ degrees of freedom. In the case where $\mathbf{x}$ was not used in the estimation of $\mathbf{\Sigma}$, $\nu=N_\lambda$.} This is the primary motivation for {\GSPICE}.

% ========================================================
% ALGORITHM:     GSPICE_gp_interp
% ========================================================
\begin{algorithm}[t]
\textbf{Inputs:} \\
$\Sigma$ = scaled covariance matrix \\
$X$ = scaled data, $N_\lambda \times N_s $ \\
$N_{\rm guard}$ = width of guard window, default 20 \\
\KwResult{prediction and variance for every pixel of every object.}
\vspace{0.1cm}
%\textbf{Notes:} .... 
\vspace{0.2cm}
\SetAlgoLined
 
 \For{$i$ = 1 to $N_\lambda$}{
    $k_* \leftarrow i$ \;
    $k \leftarrow$ ind of pixels with $|i-k| > N_{\rm guard}$ \;
    $\tilde{X}_{i,:}, \tilde{\sigma}_{i}^2 \leftarrow\  $GSPICE\_gaussian\_estimate$(k, k_*, \Sigma, X$)
 }
 return  $\tilde{X},\tilde{\sigma}^2$
 \caption{\texttt{GSPICE\_gp\_interp}}
 \label{alg:gp_interp}
\end{algorithm}

We identify ``outlier pixels'' (Alg.~\ref{alg:gp_interp}) by looping over pixels in a spectrum and estimating the mean \textit{and variance} of each (Eqs. (\ref{eq:condmean2}), (\ref{eq:condvar})), conditional on all the others except for a ``guard window'' near the predicted pixel (see Figure \ref{fig:nguard}). We use the guard window to leverage the physical correlations in the spectra, not those introduced by the instrumental LSF or upstream processing.
%In the case of LAMOST data (see \S \ref{sec:lamost}), the continuum normalization smooths the spectrum by 20 pixels and then divides by the smoothed spectrum. This imposes a correlation with neighboring pixels within the smoothing window.  We mask $N_{\rm guard}$ pixels either side of the pixel of interest (Figure \ref{fig:nguard}), with $N_{\rm guard}=20$ for LAMOST. 
%In this way, we compute the Gaussian estimate of the mean and variance for each pixel, and compare to the data. 

%\subsubsection{Case 2: Blockwise estimation}

With the mean and variance of each pixel in hand, we compute the $Z$-score for every pixel $i$ of spectrum $s$,
\begin{equation}
     Z_{is} \equiv\frac{\rm data-prediction}{\rm uncertainty}
      = \frac{{X}_{is} - \tilde{X}_{is}}{\tilde{\sigma}_{i}}\ .
\end{equation}
The criterion for determining a pixel is bad enough to replace is application dependent. There is a natural tension between replacing too few (leaving artifacts that could easily have been repaired) and too many (possibly destroying information, e.g., about rare spectral features and unusual metal abundances).

In addition to the ability to identify outliers, corrupted data, etc., based on an estimated sample covariance matrix, we might also be concerned that our estimate of the sample covariance itself (which could be based on corrupted data) may have reduced ability to identify outliers and corrupted data. Even if the spectra used to estimate $\mathbf{\Sigma}$ are selected to be high SNR and generally clean, it is still often necessary to iteratively mask the data by first masking outliers with a high threshold (say, 20$\sigma$), rebuilding the covariance, and reducing this threshold at each iteration. Algorithm \ref{alg:itermask} provides an example of iterative masking of data $X$ to produce a cleaner $\Sigma$. Because neighbors of bad pixels are suspect, the $Z$-based mask is also dilated by one pixel in the $\lambda$ direction at each iteration. 

% ========================================================
% ALGORITHM:     GSPICE_covar_iter_mask
% ========================================================
\begin{algorithm}[t]
\textbf{Inputs:} \\
$X$   = scaled data, $N_\lambda \times N_s $ \\
$M_L$ = initial mask \\
$T$   = $Z$-score threshold array \\
\KwResult{prediction and $Z$ score for every pixel of every object.}
\vspace{0.1cm}
\textit{Notes: By default uses $N_{iter}=3$ and $T=[20,8,6]$.
interpolate\_mask() replaces masked pixels with a linear interpolation in the $\lambda$ direction.  dilate() expands the mask by 1 pixel in the $\lambda$ direction.}
\vspace{0.2cm}
\\
\SetAlgoLined
 compute number of masked pixels for each spectrum \;
 discard spectra with $>64$ bad pixels \;
 initialize mask $M = M_L$ \;
 \For{$n$ = 1 to $N_{iter}$}{
    $X_M \leftarrow $ interpolate\_mask($X,M$) \;
    $\Sigma \leftarrow $ gspice\_covar($X_M$) \;
    $\tilde{X},\tilde{\sigma}^2$ = gspice\_gp\_interp($\Sigma, X, N_{\rm guard}$) \;
    $Z_{is} \leftarrow  (X_{is}-\tilde{X}_{is})/\tilde{\sigma}_{i}$\;

    $M_Z \leftarrow |Z| > T[n]$ \;
    $M_Z \leftarrow$ dilate($M_Z$, 1) \;
    $M \leftarrow M$ or $M_Z$ \;
 }
 return  $M$
 \caption{\texttt{GSPICE\_covar\_iter\_mask}}
 \label{alg:itermask}
\end{algorithm}

We now turn to outlier repair, i.e., substitution of an estimated mean and variance for each masked pixel. Once a clean covariance has been obtained, perhaps from a subset of high-quality spectra, the interpolation procedure (Alg.~\ref{alg:gp_interp}) is run on all spectra, and a $Z$-score is produced. The mask may be determined by a simple threshold. In some cases, it may be appropriate to only replace pixels above one threshold (say, 10$\sigma$), conditioned on pixels below a second threshold (say, 5$\sigma$).  That is, some pixels ($5< |Z| <10$) are neither bad enough to discard, nor good enough to trust as a reference.  The substituted values are computed with a variant of Alg.~\ref{alg:gce} that interpolates all $k_*$ pixels for a single spectrum inside the loop, rather than interpolating a single pixel for all spectra. The loops are actually done with matrix multiplication, and routines are identical except for the order of the multiplication (for speed).

%%%%%%%%%%%%%%%%%%%%%%%%%%%%%%%%%%%%%%%%%%%%%%%%%%%%%%%%
%%%%%%%%%%%%%%%%%%%%  SECTION 3  %%%%%%%%%%%%%%%%%%%%%%%
%%%%%%%%%%%%%%%%%%%%%%%%%%%%%%%%%%%%%%%%%%%%%%%%%%%%%%%%

\section{Identifying Instrumental Artifacts with LAMOST}
\label{sec:lamost}

%==================================================================
%=  FIGURE Stellar Parameters   fig:params
%==================================================================
\begin{figure}[t!]
\begin{center}
\includegraphics[width=3.35in]{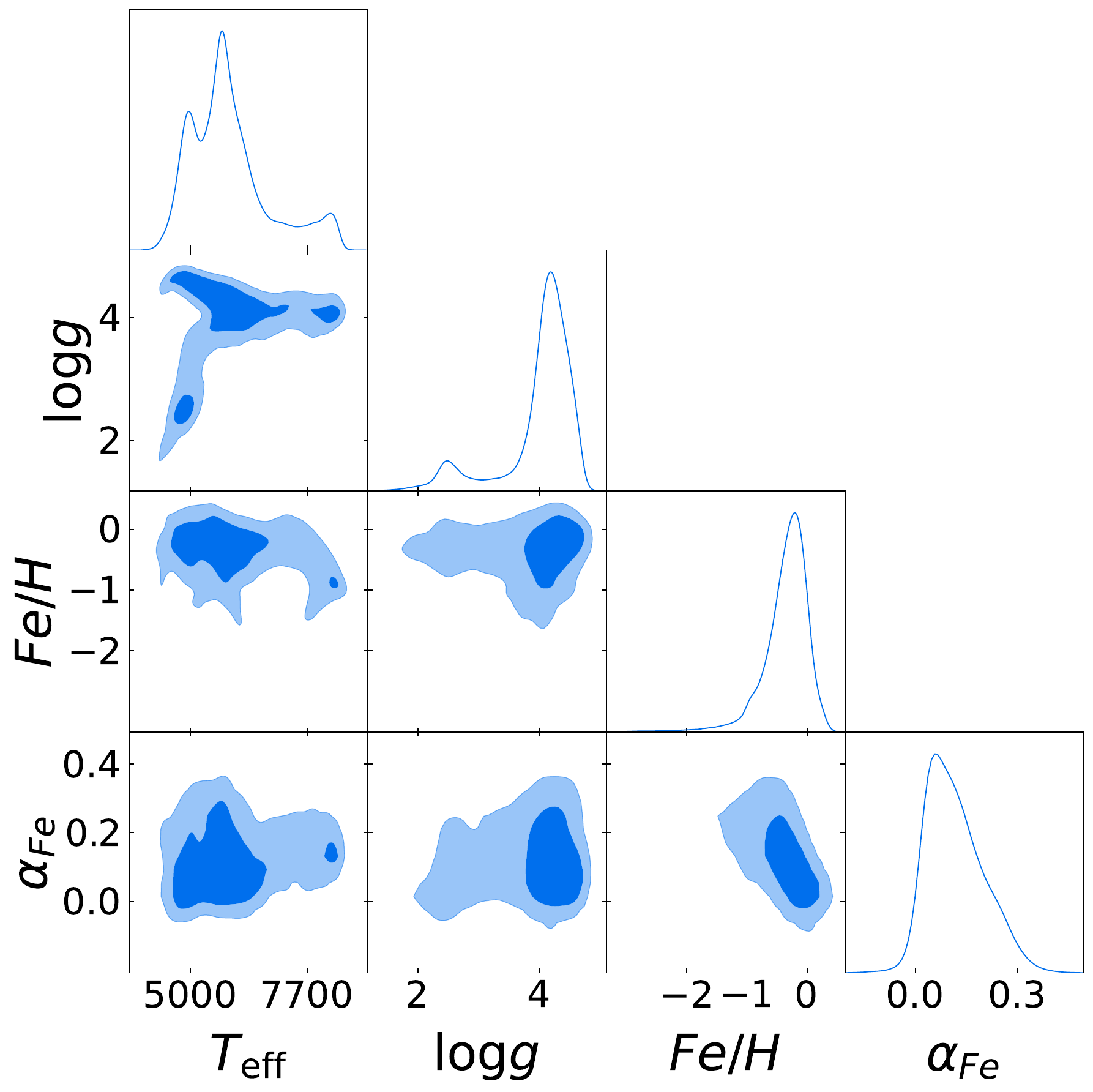}
\end{center}
\caption{A corner plot showing the estimated distribution of the stellar parameters in LAMOST DR5 that are used to compute the covariance matrix shown in Figure \ref{fig:cov} and for the experiments highlighted in later figures. This highlights that while the underlying stellar population is heterogeneous with a very non-Gaussian distribution, the conditional Gaussian estimation employed by {\GSPICE} still remains valid.}
\label{fig:params}
\end{figure}

%==================================================================
%=  FIGURE Block GSPICE   fig:blockgspice
%==================================================================
\begin{figure}[t!]
\begin{center}
\includegraphics[width=3.4in]{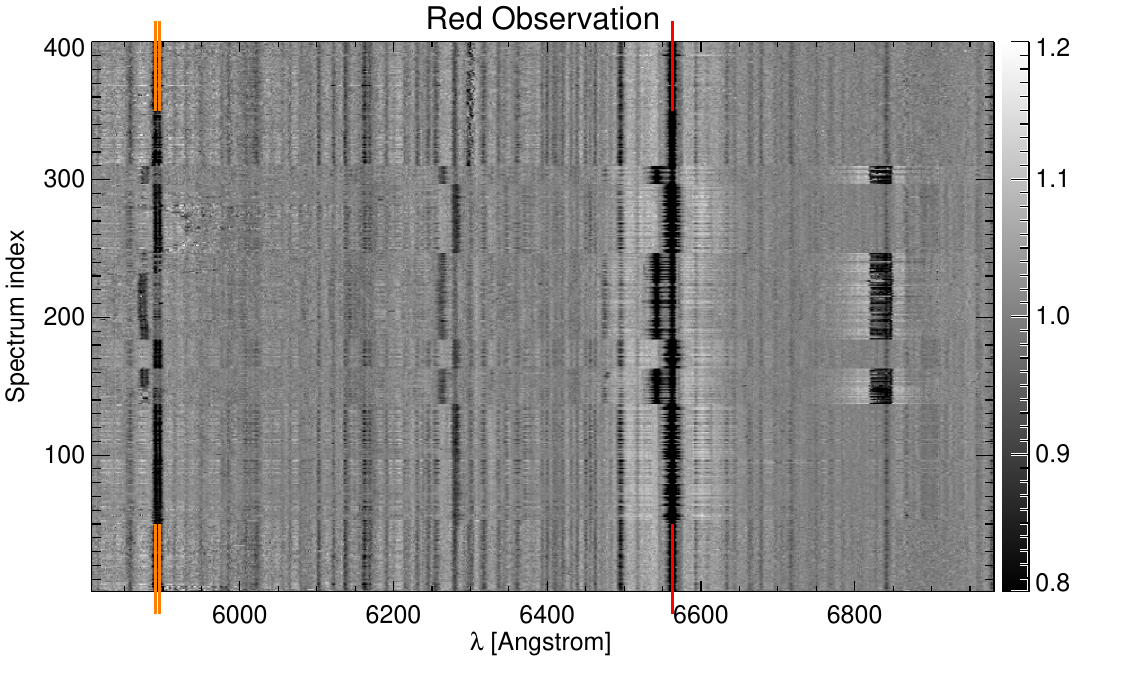}
\includegraphics[width=3.4in]{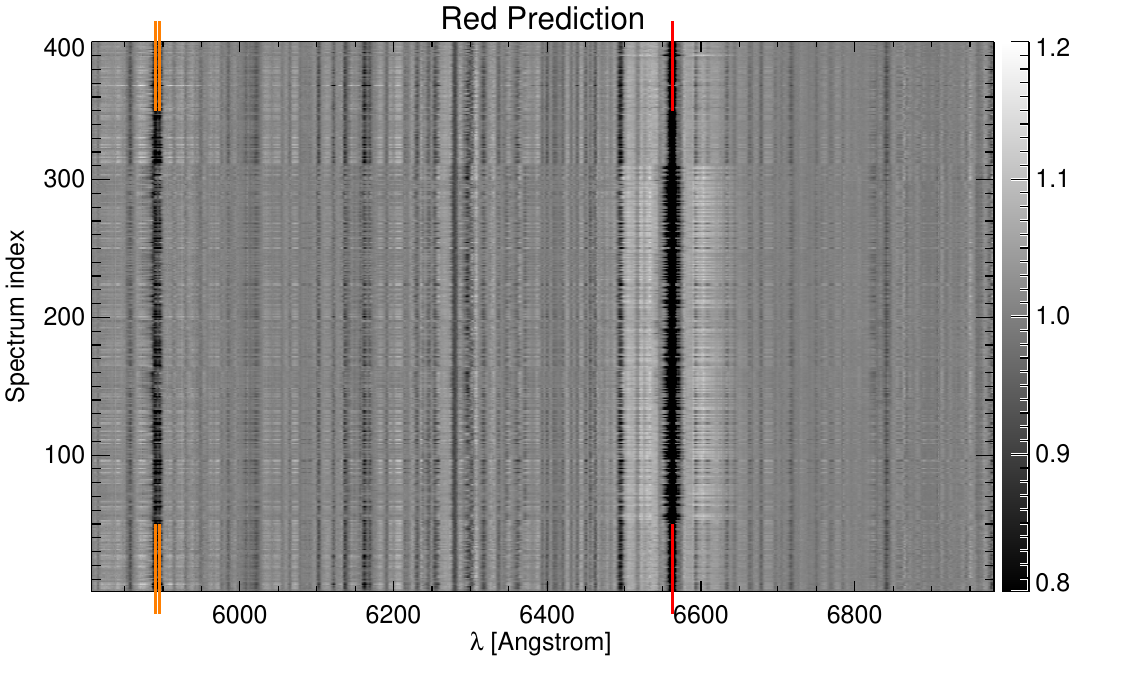}
\includegraphics[width=3.4in]{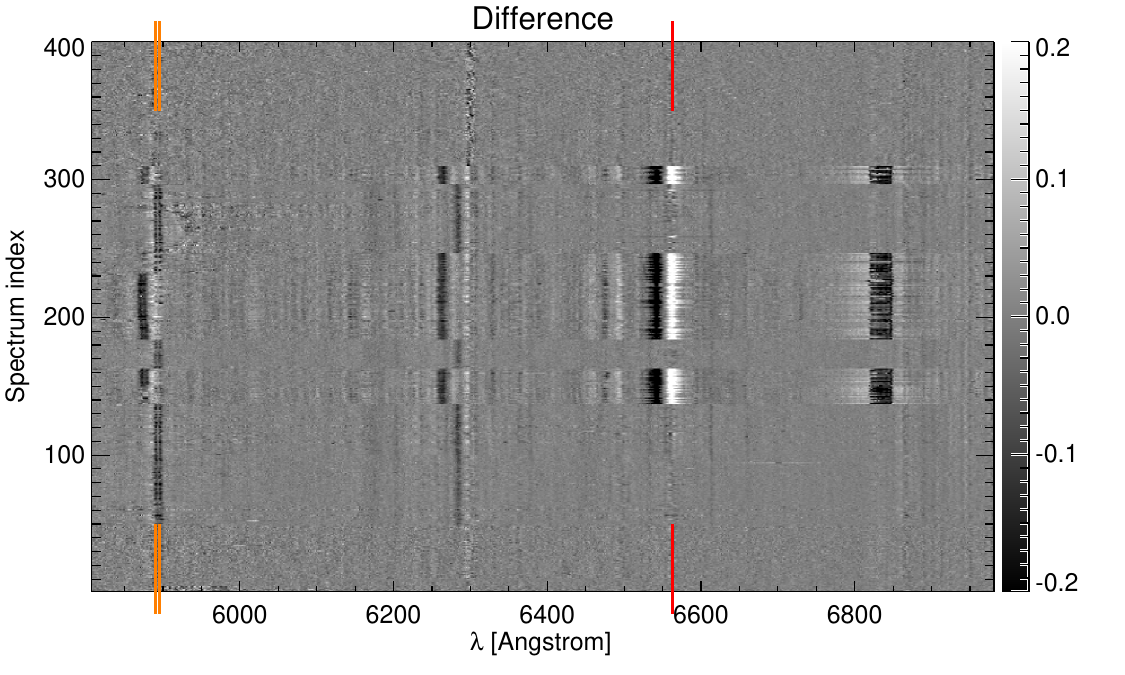}
\end{center}
\caption{An example of blockwise {\GSPICE}, predicting the red side of the spectrum (only part of which is shown) from the blue side ($\lambda < 5742$\AA). Some observed spectra (\textit{upper}) contain lines shifted relative to the {\GSPICE} prediction (\textit{middle})  due to a wavelength calibration error. This discrepancy is especially apparent in the difference image (\textit{lower}). {\GSPICE} predicts the mean and variance of each pixel, so non-conforming pixels may be found with a simple $Z$-score cut. This wavelength problem does not appear in LAMOST DR4, but affects $\sim 1\%$ of the spectra in an unreleased version of DR5 (M. Xiang, priv. comm). The wavelengths of the Na D doublet (\textit{orange}) and H$\alpha$ (\textit{red}) are indicated by vertical lines.}
\label{fig:blockgspice}
\end{figure}

%\textbf{TODO: Rewrite to provide a better overview and merge text into the previous section. Add overview plot of LAMOST sample properties?}

We demonstrate the functionality of {\GSPICE} by applying it to data from the Large Sky Area Multi-Object Fiber Spectroscopic Telescope (LAMOST) \citep{LAMOST:2015}. The Gaussian estimation machinery can be used in two modes: pixelwise and blockwise. These are described below:
\begin{itemize}
    \item In the \textbf{pixelwise} case discussed in Section \ref{sec:method}, small artifacts (bad columns, cosmic ray hits, etc.) can be detected with greater sensitivity. To find subtle defects, it is helpful to use as much spectral range as possible as the basis for prediction for each test pixel, and then loop over test pixels. This requires more computation, but is often tractable. 
    \item In some cases, hardware or software problems may render a large fraction of the spectrum corrupted. The pixelwise estimate will be corrupted if too many of the reference pixels are bad. An example of \textbf{blockwise} estimation (also discussed in Section \ref{sec:method} but in less detail) is estimating the red half of the spectrum from the blue half, or vice versa. This makes the estimate on a corrupted half of the spectrum reliable if it is conditioned on the good half. A drawback is that because only half of the data is used as a basis for prediction, the prediction is somewhat noisier. 
\end{itemize}
Blockwise and pixelwise estimation can be done per star, respecting a pre-existing bad-pixel mask, or they can be applied to all stars at once. 

%==================================================================
%=  FIGURE One star   fig:onestar
%==================================================================
\begin{figure*}[t!]
\begin{center}
\includegraphics[width=7.5in]{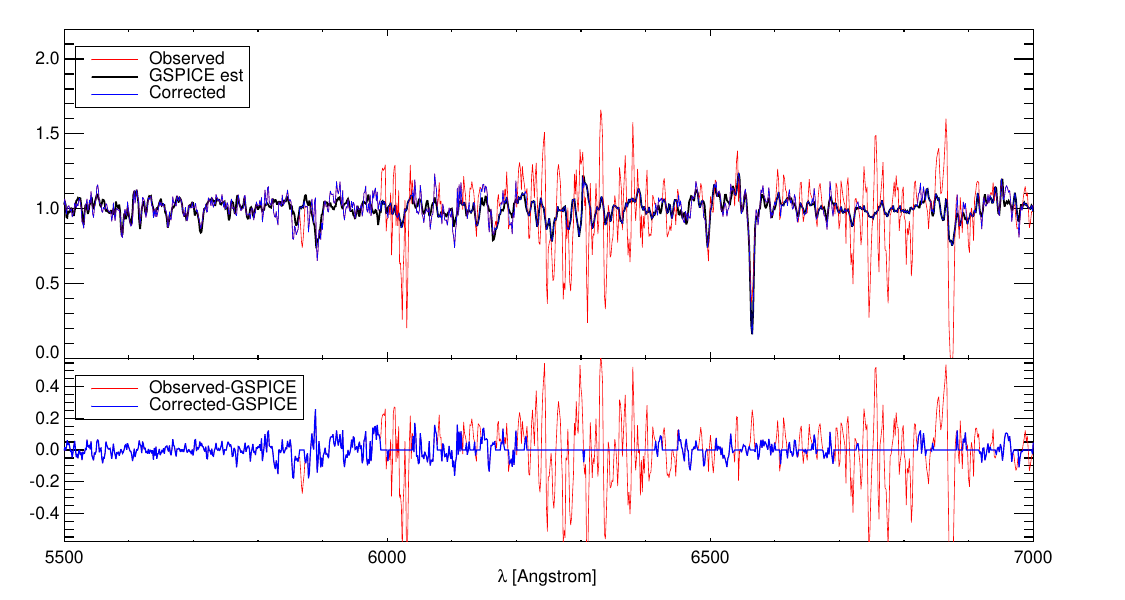}
\end{center}
\caption{\textit{Upper panel:} {\GSPICE} repair of a single star (LAMOST ID 20111207-B90305-12-078 at $\alpha=140.65498, \delta=32.71338$).  The observed continuum-normalized spectrum (\textit{red}) exhibits unphysical oscillations at $\lambda > 6000$\AA. In pixels where these disagree significantly with the {\GSPICE} estimate (\textit{black}), they are replaced by the estimate in the corrected spectrum (\textit{blue}). \textit{Lower panel:} Observed spectrum minus {\GSPICE} prediction (\textit{red}) and corrected minus {\GSPICE} prediction (\textit{blue}).}
\label{fig:onestar}
\end{figure*}

%==================================================================
%=  FIGURE Repair   fig:repair
%==================================================================
\begin{figure*}[t!]
\begin{center}
\includegraphics[width=3.5in]{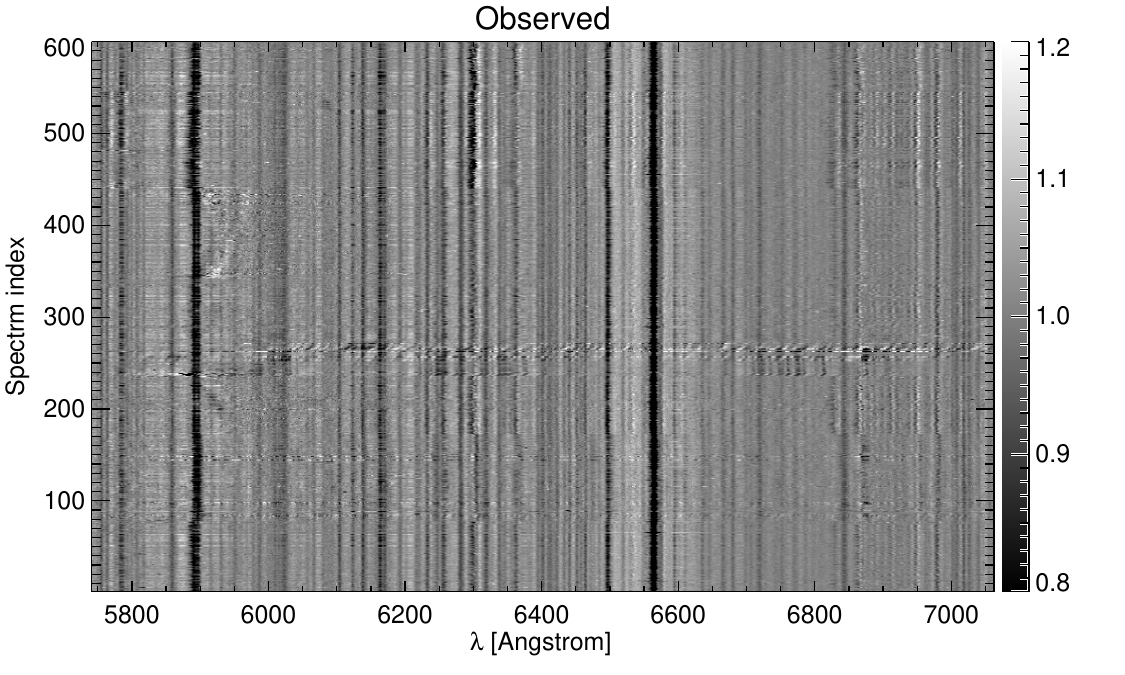}
\includegraphics[width=3.5in]{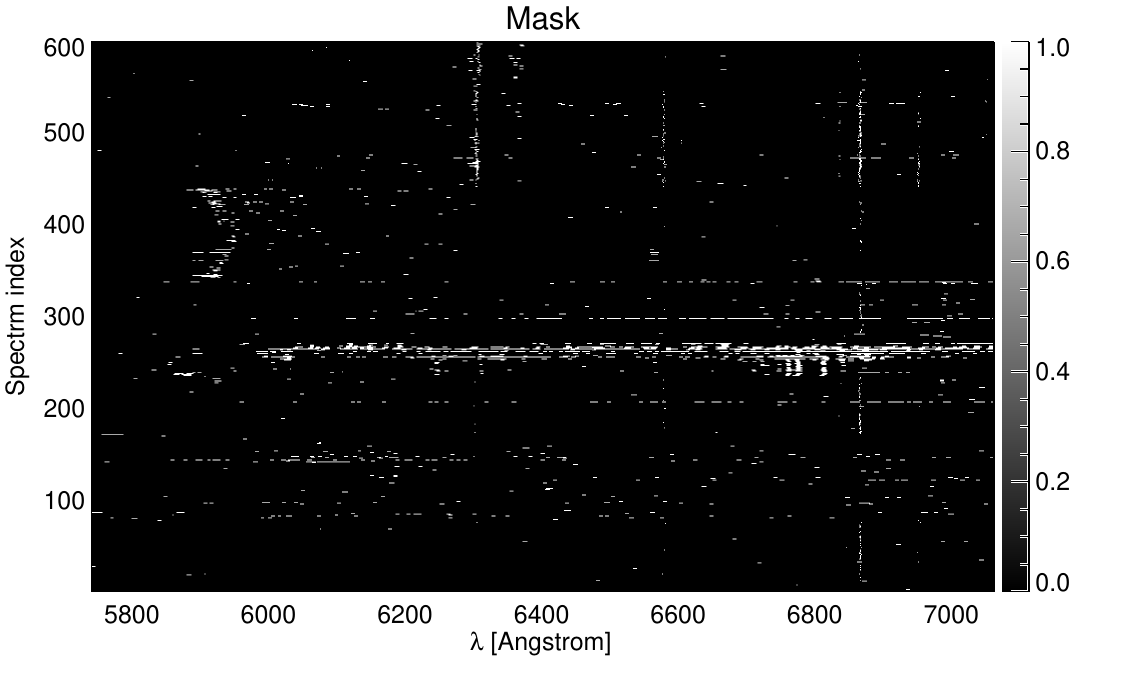}
\includegraphics[width=3.5in]{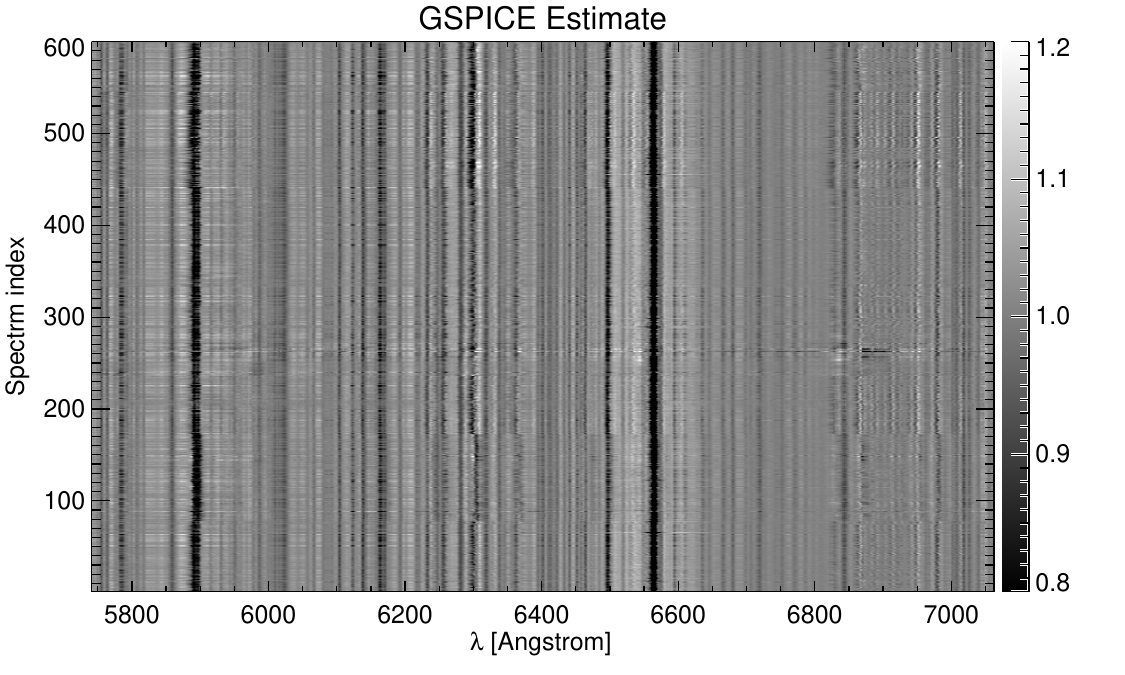}
\includegraphics[width=3.5in]{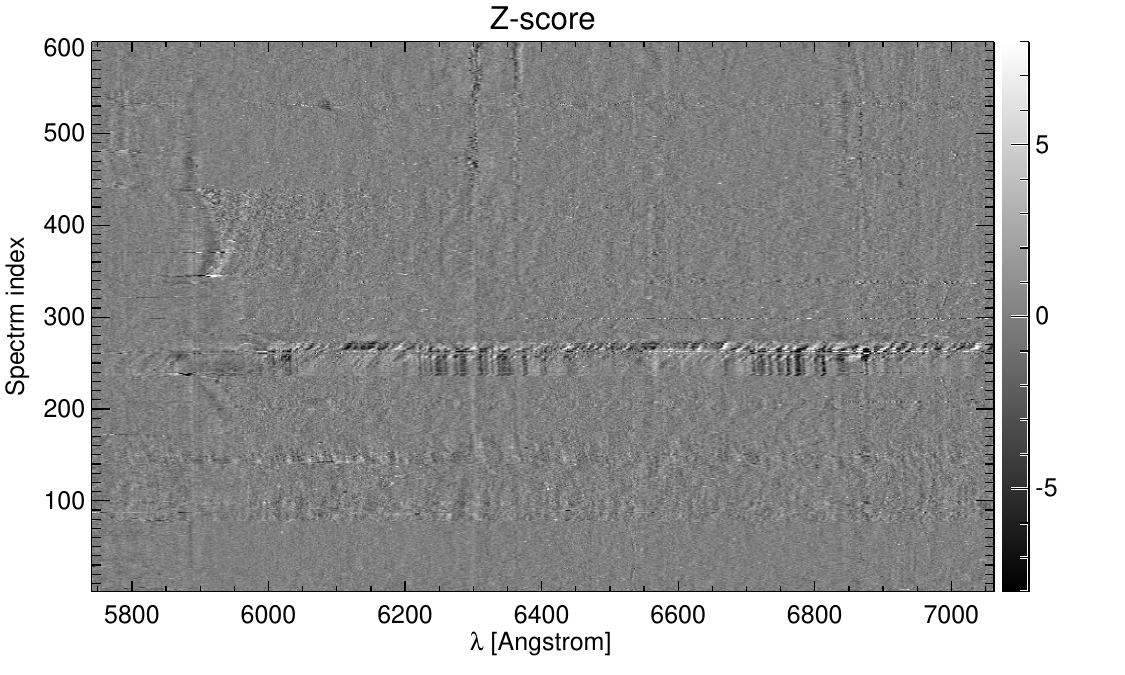}

\includegraphics[width=3.5in]{figs/repair-msk.pdf}
\includegraphics[width=3.5in]{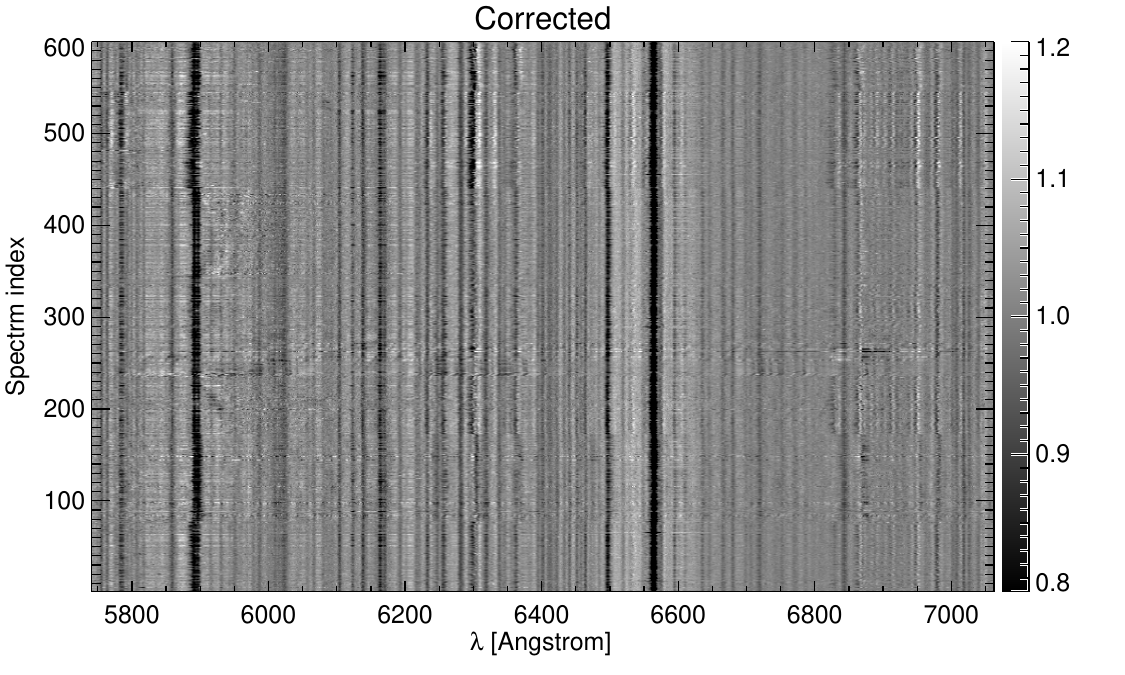}
\end{center}
\caption{Repair of artifacts in LAMOST data. {\GSPICE} is applied to the observed data (\textit{upper left}) to estimate the mean and variance. These are used to evaluate $Z=$(Observed-Predicted)/$\sqrt{\rm var}$ (\textit{upper right}). Pixels with $|Z|>10$ are masked, and the mask is dilated by 2 pixels in the wavelength direction (\textit{lower left}).  Finally, masked pixels are replaced with the {\GSPICE} prediction (\textit{lower right}). The subtle features in the $Z$ image are below threshold for masking, but may reveal something interesting about the data processing.}
\label{fig:repair}
\end{figure*}

In Section \ref{subsec:lamost}, we describe the LAMOST data used in this case study. In Section \ref{subsec:blockwise}, we highlight {\GSPICE}'s ability to identify and correct large-scale data artifacts identified by our blockwise estimator. In Section \ref{subsec:pixelwise}, we do the same for smaller, local regions identified by our pixelwise estimator.

\subsection{LAMOST Data}
\label{subsec:lamost}

%\textbf{TODO: Move spectral covariance here. Clean up and expand language a bit.}

The LAMOST telescope has an effective aperture of 4m with a $5^\circ$ field of view.  The instrument is comprised of 16 spectrographs fed by 250 fibers each, for a total of 4000 fibers with a 3.3 arcsecond diameter.  In each spectrograph, light passes through a dichroic element to a red camera and a blue camera, each with a 4k$\times$4k CCD.  The red and blue sides of the spectrum are combined into a composite spectrum with wavelength bins uniform in $\log\lambda$.  The bin width corresponds to $\Delta\log\lambda=10^{-4}$, or about $69$ km/s.  The fifth data release\footnote{Available at http://dr5.lamost.org} contains 9,026,365 spectra with a limiting magnitude of $r=19$ at resolution 1800, covering a wavelength range from 3750-9000 \AA.  Most stars are dwarfs with temperatures between 4500 and 7000K, and metallicities between -1 and 0 (Fig. \ref{fig:params}). 

We use a sample of 3.9 million spectra with $S/N > 30$ in i-band. While LAMOST provides quality flags, we choose to ignore them all since our goal is to demonstrate what would happen in the case of unexpected artifacts.  We do, however, exclude spectra with a large number of masked pixels from the computation of the covariance. For LAMOST DR5, $N_\lambda=3801$.

\subsection{Blockwise \texttt{GSPICE} for Large-Scale Artifacts}
\label{subsec:blockwise}

%\textbf{TODO: Clean up text. Replace figure with modified version gspice5.}

Blockwise use of {\GSPICE} is convenient for finding problems that affect a broad wavelength range, but are assumed to be isolated from some other wavelength range. In the following example, we apply Eqs. (\ref{eq:condmean2}) and (\ref{eq:condvar}) to the LAMOST sample, with the pixels to be estimated ($k_*$) corresponding to the entire red half of the spectrum and the pixels to be conditioned on ($k$) corresponding to the blue half of the spectrum. In roughly 99\% of cases, the prediction agrees broadly with the observation, but in some cases, there are striking differences. Figure \ref{fig:blockgspice} shows a set of spectra from one field, ordered by spectrograph. These composite spectra contain data from more than one exposure, and it appears that in at least one exposure, some of the spectrographs have an incorrect wavelength calibration.  

\subsection{Pixelwise \texttt{GSPICE} for Small Outlying Regions}
\label{subsec:pixelwise}

%\textbf{TODO: Clean up text. Add figure gspice4.}

Pixelwise GCE (or the default implementation of {\GSPICE}) aims to provide a more sensitive detection of localized outliers by using all information in the spectrum except that near the test pixel. That is, $k_*$ is a single pixel and $k$ is all other pixels \textit{except} those in a guard window of length $N_{\rm guard}$ (Figure \ref{fig:nguard}). This is slower than blockwise GCE, because we must loop over each pixel. An example of an application of how this approach can be applied to infilling individual spectra is shown in Figure \ref{fig:onestar}.

The thought of looping over $N_\lambda \sim 4000$ pixels in $\sim 4$ million spectra, requiring 16 billion effective matrix inversions, may be somewhat daunting.  Even if Eqs. (\ref{eq:condmean2}) and (\ref{eq:condvar}) can be evaluated in 1 ms, this is still 16 Ms or about 4000 core hours. For maximum sensitivity to outliers, this can be done. However, the following procedure is substantially more efficient:
\begin{itemize}
\item Compute the sample covariance matrix.
\item For each star, replace masked pixels with \GSPICE estimates ($N_{\rm spec}$ inversions).
\item For each pixel, estimate that pixel for all stars ($N_\lambda$ inversions).
\item Compute $Z$, mask values if $|Z| > {\rm threshold}$, and then replace with estimate.
\item Repeat from step (1) with a different threshold.
\end{itemize} 

This requires on the order of a few $\times (N_{\rm spec} + N_\lambda)$ matrix inversions, enormously faster than the alternative of $N_{\rm spec} \times N_\lambda$ matrix inversions (as $N_\lambda \ll N_{\rm spec}$). The result of this procedure may be seen in Figure \ref{fig:repair}.

%%%%%%%%%%%%%%%%%%%%%%%%%%%%%%%%%%%%%%%%%%%%%%%%%%%%%%%%
%%%%%%%%%%%%%%%%%%%%  SECTION 4  %%%%%%%%%%%%%%%%%%%%%%%
%%%%%%%%%%%%%%%%%%%%%%%%%%%%%%%%%%%%%%%%%%%%%%%%%%%%%%%%
\section{Detecting Astrophysical Interlopers with Diffuse Interstellar Bands}
\label{sec:DIB}
%\textbf{TODO: Clean up text.}

Beyond instrumental artifact detection, {\GSPICE} can also be used to detect astrophysical artifacts and interlopers. To demonstrate the use of GSPICE for anomaly detection, we make use of diffuse interstellar bands (DIBs) and atomic absorption features. DIBs are absorption features with varying widths and strengths, some well correlated with reddening, some not. Though they were recognized decades ago \citep{Merrill:1934,Herbig:1975}, and over 400 have been cataloged \citep{Hobbs:2009}, very few have been conclusively identified. They are thought to arise from CH stretching modes in small carbonaceous dust grains, and are worthy of study in their own right as a tracer of interstellar chemistry. 

Measuring ISM absorption of background starlight depends on an accurate estimate of the unabsorbed stellar spectrum. This is especially challenging for the Na D doublet, which appears prominently in a range of stellar types. High-resolution spectroscopy can separate the much narrower ISM doublet from the stellar doublet, but for more modest resolutions (e.g. SDSS, LAMOST, DESI), instrumental broadening prevents using the line width as a distinguishing feature. Fortunately, estimation of the unabsorbed spectrum near features of interest, conditional on the rest of the spectrum, is precisely what {\GSPICE} provides.\footnote{Note that this approach of exploiting high-dimensional Gaussian conditional estimation in the underlying spectrum is also applied in \citet{madgics2023}, who also use it to detect and trace DIBs. While their multi-component approach is more naturally suited to detecting these types of astrophysical signals, we nonetheless consider this a useful exercise.}

In order to detect this type of dust feature using {\GSPICE}, it is necessary to build both a ``clean'' covariance matrix and a ``dusty'' one.  The clean covariance is computed from stars with very low foreground reddening ($E_{B-V} < 0.04$) as measured by \cite{Schlegel:1998}.  This covariance is used to establish the unabsorbed spectrum.  An estimate based on the clean covariance does not ``know'' about ISM absorption, so it cannot use weak unmasked DIBs to corrupt the estimate.  However, such a covariance would be inappropriate for outlier identification because it would flag genuine absorption features as outliers.  We therefore build a dusty covariance matrix with no cut on reddening.\footnote{For the moment, we neglect the fact that these two estimates of the covariance may be constructed with rather different stellar types.}  The variance of pixels is higher near these features, and they covary appropriately.  After a pass of {\GSPICE} to flag outliers, we use the clean covariance to predict each stellar spectrum in the regions of interest.  Subtracting the predicted spectrum from the observation yields the measurement. 

%==================================================================
%=  FIGURE DIGS   fig:dibs
%==================================================================
\begin{figure}[t!]
\begin{center}
\includegraphics[width=3.5in]{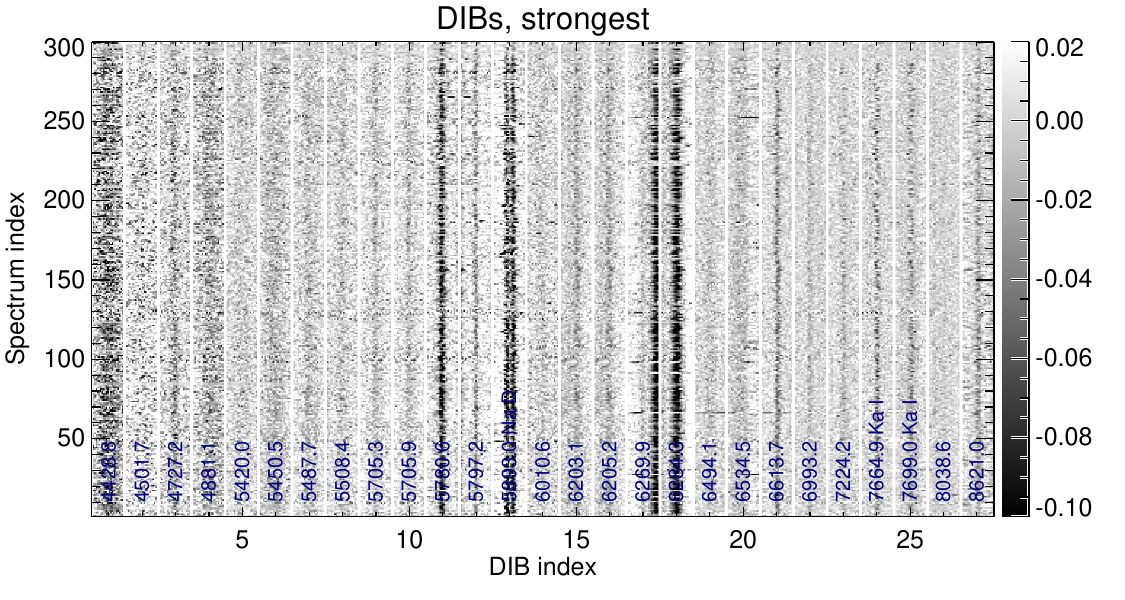}
\includegraphics[width=3.5in]{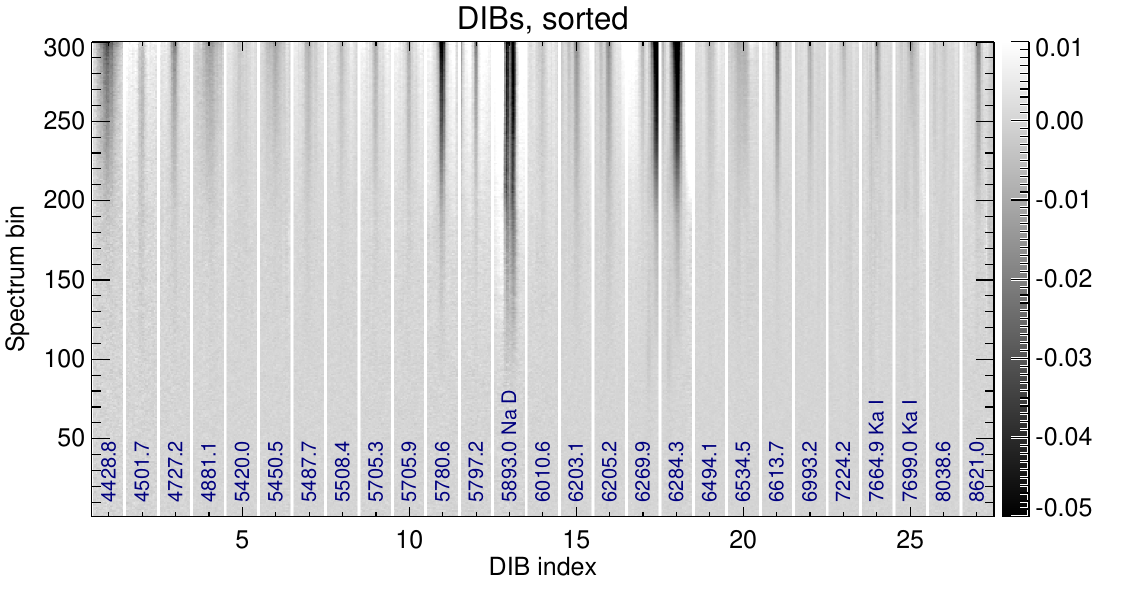}
\end{center}
\caption{Reconstructed DIB absorption (observed spectrum minus predicted clean spectrum) in 21-pixel regions centered on 24 DIBs (i.e., $|\Delta \log\lambda| < 0.001$).  The Na D doublet and two K{\textsc i} lines are included for reference.  Labels are air wavelengths in \AA ngstroms from \cite{Hobbs:2009}. These are multiplied by 1.00028 to convert to the vacuum wavelengths used by LAMOST. \textit{Upper panel:} absorption features for the 300 stars with the most foreground dust. \textit{Lower panel:} 192,000 arbitrarily chosen stars sorted by foreground dust (each row is an average of 640 stars). White halos around dark lines are due to the LAMOST continuum normalization.  The grayscale represents fractional absorption.}
\label{fig:dibs}
\end{figure}

%==================================================================
%=  FIGURE DIB scatter   fig:dibscatter
%==================================================================
\begin{figure*}[t!]
\begin{center}
\includegraphics[width=4.8in]{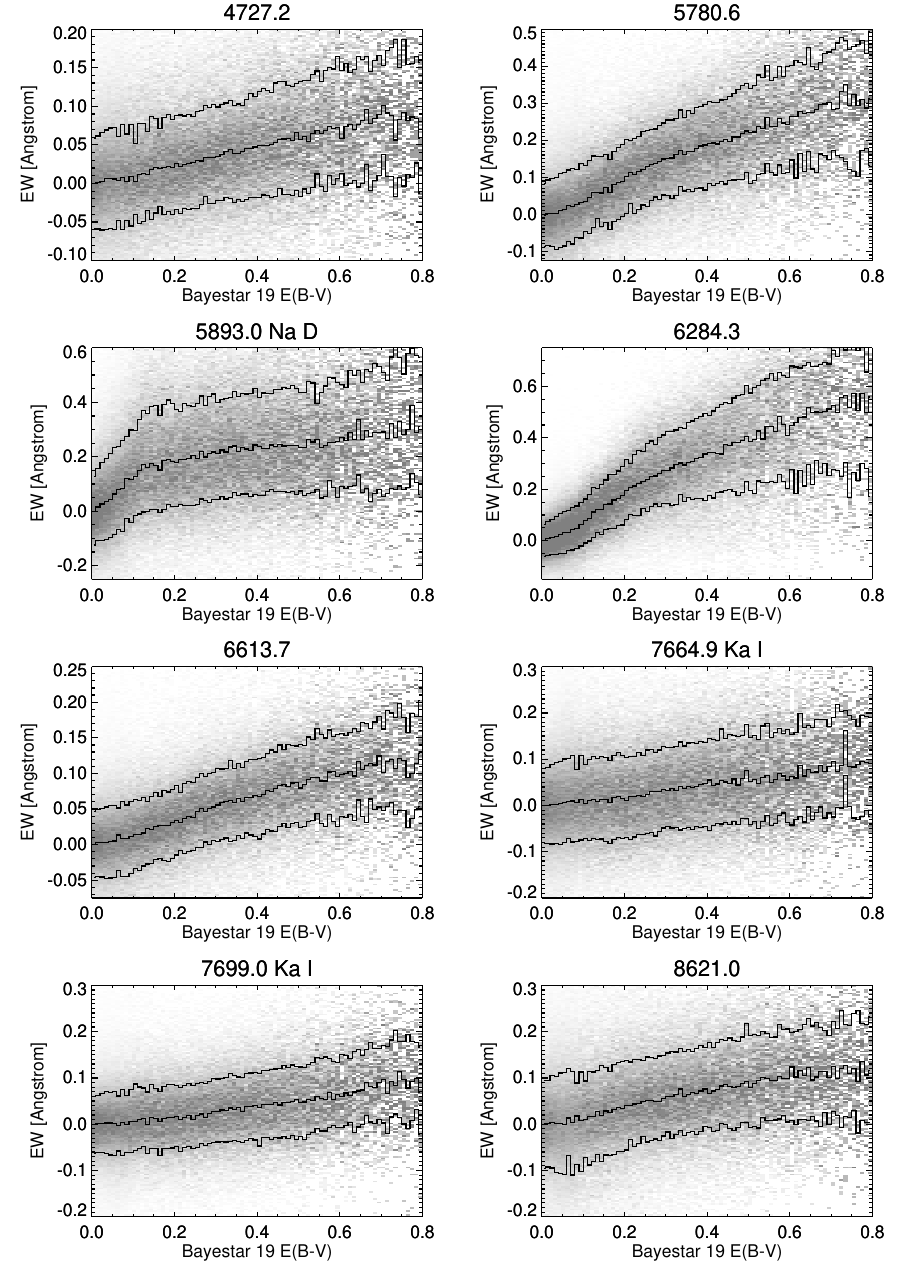}
\end{center}
\caption{Absorption line equivalent width in a 7-pixel window vs. Bayestar19 $E(B-V)$ for selected DIBs and atomic lines for the same spectrum sample as in Fig. \ref{fig:dibs}.  Some DIB features appear to be quite linear with dust column density (4727.2\AA), while others are quite non-linear (5780.6\AA \ \& 6284.3\AA). The Na doublet saturates above a reddening of approximately 0.1 mag due to self absorption. }
\label{fig:dibscatter}
\end{figure*}

We select wavelength regions within 10 pixels of the 20 strongest DIBs in the LAMOST spectral range from the \cite{Hobbs:2009} catalog, and also include the Na D doublet ($\lambda 5890,5896$) and the K{\sc i} $\lambda7666$ and $\lambda7700$ lines.   As a measure of foreground absorption, we sample from the Bayestar19 3-D dust map \citep{Bayestar:2019} at distances drawn from the \textit{Gaia} parallax \citep{GaiaDR2:2018,GaiaDR2Parallax:2018}, using 100 samples to determine the value and variance.  Most of the absorption spectra for the 300 stars with highest foreground dust exhibit absorption features (Figure \ref{fig:dibs}). It is evident that there is some non-linearity; for selected DIBs and Na D, the dependence on reddening is linear at low column and then begins to saturate (Figure \ref{fig:dibscatter}).

While the motivation for {\GSPICE} is not to measure DIBs, the DIB analysis nonetheless shows the power of a Gaussian prior on spectrum space for estimating deviations from that prior (in this case DIB absorption). The LAMOST resolution of $\sim 150$ km/s (with 69 km/s pixels) is not ideal for getting dust radial velocities as in \cite{Tchernyshyov:2018}, but this approach could be valuable for such work, or any other time that an unabsorbed continuum is needed. In the same vein, Na D estimates for the \textit{stars} are probably also more reliable with this technique, though the estimates undoubtedly leverage the presence of other metals in the spectrum.  Alternatively, one could rely on Bayestar19 and a sufficiently sophisticated model of interstellar absorption to make the correction.  Both approaches could be applied to large spectroscopic samples, but they rely on assumptions that must be carefully checked with high-resolution spectroscopy. 

% \subsection{Assessing results vs. stellar type}
% We aspire to look at reconstruction quality vs. $T_{eff}$, $\log g$, Fe/H, etc.  

%%%%%%%%%%%%%%%%%%%%%%%%%%%%%%%%%%%%%%%%%%%%%%%%%%%%%%%%
%%%%%%%%%%%%%%%%%%%%  SECTION 5  %%%%%%%%%%%%%%%%%%%%%%%
%%%%%%%%%%%%%%%%%%%%%%%%%%%%%%%%%%%%%%%%%%%%%%%%%%%%%%%%
\section{Discussion and Conclusions}
\label{sec:concl}

%\textbf{TODO: Broaden discussion a bit. Wrap up conclusions. Mirror introduction and abstract.}

Motivated by the need to find outliers in spectral data, we have applied the concept of Gaussian conditional estimation in novel ways.  By thinking of stellar spectra as generated by a (non-stationary) Gaussian process, sampled at $N_\lambda$ wavelengths, we can ask if a particular spectrum conforms to a prior expressed by a covariance matrix. Astrophysical deviations, as well as those caused by instrumental defects and software bugs, can readily be detected. Importantly, ``unknown unknowns'' may be revealed, because the expectation is not crafted with expert knowledge, but rather specified by a training data set. This data-driven approach has risks and benefits in common with other data-driven/machine learning approaches. 

The novelty of our proposed approach (\GSPICE) is the inherently \textit{pixelwise} nature of the estimate, which loops over pixels and predicts each conditional on the rest of the spectrum. This approach finds localized defects with impressive sensitivity, leveraging both local behavior in conjunction with global structure. If some spectral regions are known to be suspect (e.g near ISM absorption features) they can be predicted conditional on the rest of the spectrum. Importantly, due to the conditional nature of the estimate and the large number of pixels we can use to condition the estimate on, we can still construct excellent pixelwise (and even blockwise) estimates even while the joint distribution of sources is clearly poorly jointly modeled by a multivariate Gaussian distribution.

The {\GSPICE} approach may be applied in any vector space where a Gaussian prior is adequate. For example, hyperspectral imaging (a 3-D dataset with a spectrum of every point in a 2-D image) has a similar need for artifact masking and anomaly detection. The adaptive coherence estimator \citep[ACE;][]{ACE:2005} is commonly employed for detection of a spectral signature (e.g., methane over an oil field). ACE requires signature-free training data to build a sample covariance matrix that can be used to extract the desired signature in the presence of clutter and interference. The covariance is intended to express ``normal'' variation in the spectral, with the desired signature (or perhaps an unknown one) detectable as an outlier. Single-pixel outliers in the spectra can degrade the covariance estimate, reducing sensitivity to the signatures of interest. {\GSPICE} might offer a path to improved covariance estimates for use with ACE. 

{\GSPICE} could also be applied to images. Astronomers usually estimate a point-spread function from small images of stars, appropriately centred and normalized. The prior on the PSF and its variation due to atmospheric variation, mis-centering, etc., could be expressed as the covariance of flux in pixel $i$ with pixel $j$. Hot pixels could be detected and repaired with {\GSPICE}, yielding a clean representation of the PSF. The advantage of this representation is that the photometric measurement can marginalize over PSF variation analytically. Works such as \citet{2022ApJ...933..155S} that have applied these types of methods in those domains have already found good success.

Additionally, while in this paper we showcase how stellar spectra can be modeled with {\GSPICE}, in principle, one can extend the same approach to work with spectra of galaxies. This is especially promising for ongoing spectroscopic surveys such as DESI, which will provide millions of galactic and quasar spectra. The opportunities in extending such analyses to galaxies are obvious; a data-driven approach to building a galactic spectral template is desirable, as first-principle approaches may not pick up all the nuances of real galaxies, especially those that are at high-redshift. Thus, {\GSPICE} could greatly contribute to either outlier detection, such as spectroscopically lensed galaxies, or to provide robust redshift-dependent spectral templates for hydrodynamical simulations.

The immediate challenge in extending {\GSPICE} to work with galaxies is that of redshift; for a large redshift range, building a conditional covariance matrix is non-trivial, as galaxies with low redshift do not have data on the redder end of spectra, while galaxies with high redshift do not have data on the bluer end. However, this issue can likely be solved through an iterative procedure that can be used to build redshift-dependent covariance matrices. We hope to explore this problem in a future paper.

%Some generalizations beyond the scope of this work:
%\begin{itemize}
%\item Can we measure radial velocity by preparing many covariance matrices corresponding to different velocities shifts, then evaluating $\chi^2$ for each?
%\item Can we apply this technique to continuum-subtracted spectra rather than continuum-normalized?  Variations in sensitivity across the spectral range would be represented in the training data and effectively ignored. 
%\item Can the covariance matrix be supplemented with ancillary data (colors, magnitudes, reddening estimates, abundances) and then used to predict those quantities?  The methodology may seem too linear for that, but one could input a combination of data, data$^2$, data$^3$, ...
%\end{itemize}
%This last is probably better done with an inherently non-linear approach (neural networks) but for artifact detection and repair, GSPICE is easy to understand and computationally straightforward. 

\noindent \textit{Acknowledgments:}
DPF and JSS would like to thank the Max-Planck-Institut f\"ur Astronomie for hosting them during a period of time where the majority of this work was completed. JSS (and his back) would also like to thank Rebecca Bleich for making sure he got a good desk chair.

JSS was supported by funding from the Dunlap Institute, an NSERC Banting Postdoctoral Fellowship, NSERC Discovery Grant RGPIN-2023-04849, and a University of Toronto Connaught New Researcher Award. TK was supported by the National Science Foundation Graduate Research Fellowship under Grant No. DGE - 1745303 as well as a University of Toronto Arts \& Science Postdoctoral Fellowship during a portion of time when this work was being completed. JSS and TK are also grateful for support from the Dunlap Institute. The Dunlap Institute is funded through an endowment established by the David Dunlap family and the University of Toronto.

\noindent \textit{Data:} The Guoshoujing Telescope (the Large Sky Area Multi-Object Fiber Spectroscopic Telescope LAMOST) is a National Major Scientific Project built by the Chinese Academy of Sciences. Funding for the project has been provided by the National Development and Reform Commission. LAMOST is operated and managed by the National Astronomical Observatories, Chinese Academy of Sciences. This work has made use of results from the European Space Agency (ESA) space mission \textit{Gaia}, the data from which were processed by the \textit{Gaia Data Processing and Analysis Consortium} (DPAC). Funding for the DPAC has been provided by national institutions, in particular the institutions participating in the Gaia Multilateral Agreement. The \textit{Gaia} mission website is https://www.cosmos.esa.int/web/gaia.

The authors used \texttt{NumPy} \citep{numpy} to implement \GSPICE.

%\begin{thebibliography}

\bibliography{gspice}

%\end{thebibliography}

\clearpage
\appendix
\section{Computational Details}
\label{sec:comp}

This work used the Cannon\footnote{Named in honor of Annie Jump Cannon, who developed the first spectral classification system for stars.} cluster at Harvard University.  Each Cannon node has two water-cooled Intel 24-core Platinum 8268 Cascade Lake processors with 4 GB of RAM per core.  Cascade Lake cores have dual AVX-512 fused multiply-add (FMA) units\footnote{AVX-512 is the 512-bit Advanced Vector Extensions Single Instruction Multiple Data (SIMD) instruction set for x86 cores.  Registers can hold 512 bits of data, for example 8 double-precision floats.  Fused Multiply Add means that the numbers in a register may be multiplied by those in another, and added to those in a third, all on one clock cycle.  With 2 FMA units, a core can perform $8\times2\times2=32$ double-precision operations per cycle.  A clock speed of 3.5 GHz implies a theoretical maximum of 112 billion floating-point operations (Gflops) per core.}  For matrix multiplication we use CBLAS dgemm routines\footnote{\href{https://software.intel.com/en-us/mkl-developer-reference-c-cblas-gemm}{MKL C Basic Linear Algebra Subprograms (CBLAS) dgemm}} from Intel's Math Kernel Library in IDL, and the MKL version of NumPy in Python.  These both achieve about 80 GFlops or 70\% of the theoretical maximum for matrices of the size used in this work.  On a 48-core node that amounts to 3.8 TFlops (double precision).

However important matrix multiplication may be for GSPICE, matrix \textit{inversion} would dominate if not approached thoughtfully.  A covariance matrix is positive semi-definite and symmetric, allowing use of fast Cholesky routines.  Even using Cholesky decomposition, it takes roughly 1 core-second to invert a $3801\times3801$ matrix with \texttt{dpotri} and \texttt{dpotrf}\footnote{\href{http://www.netlib.org/lapack/explore-html/d1/d7a/group__double_p_ocomputational_ga9dfc04beae56a3b1c1f75eebc838c14c.html}{LAPACK dpotri}}.  Outlier finding involves inverting 3801 matrices and applying each one to $N$ spectra.  Repair involves inverting one matrix per spectrum, because each spectrum (in general) has a different set of pixels masked.  This requires millions of matrix inversions for a survey like LAMOST, and if rejection is iterated, perhaps billions. 

In this section we use the fact that we are inverting many matrices that are almost identical to achieve a speedup of 2-3 orders of magnitude.  We have the full $N\times N$ covariance matrix, $M$, and repeatedly extract a submatrix $A$ containing most of the rows and columns of $M$.  In the following we present a way to compute $A^{-1}$ given that we already have $M^{-1}$. 

\subsection{Inverse of a submatrix}
\begin{figure}[t!]
\begin{center}
\includegraphics[width=3.5in]{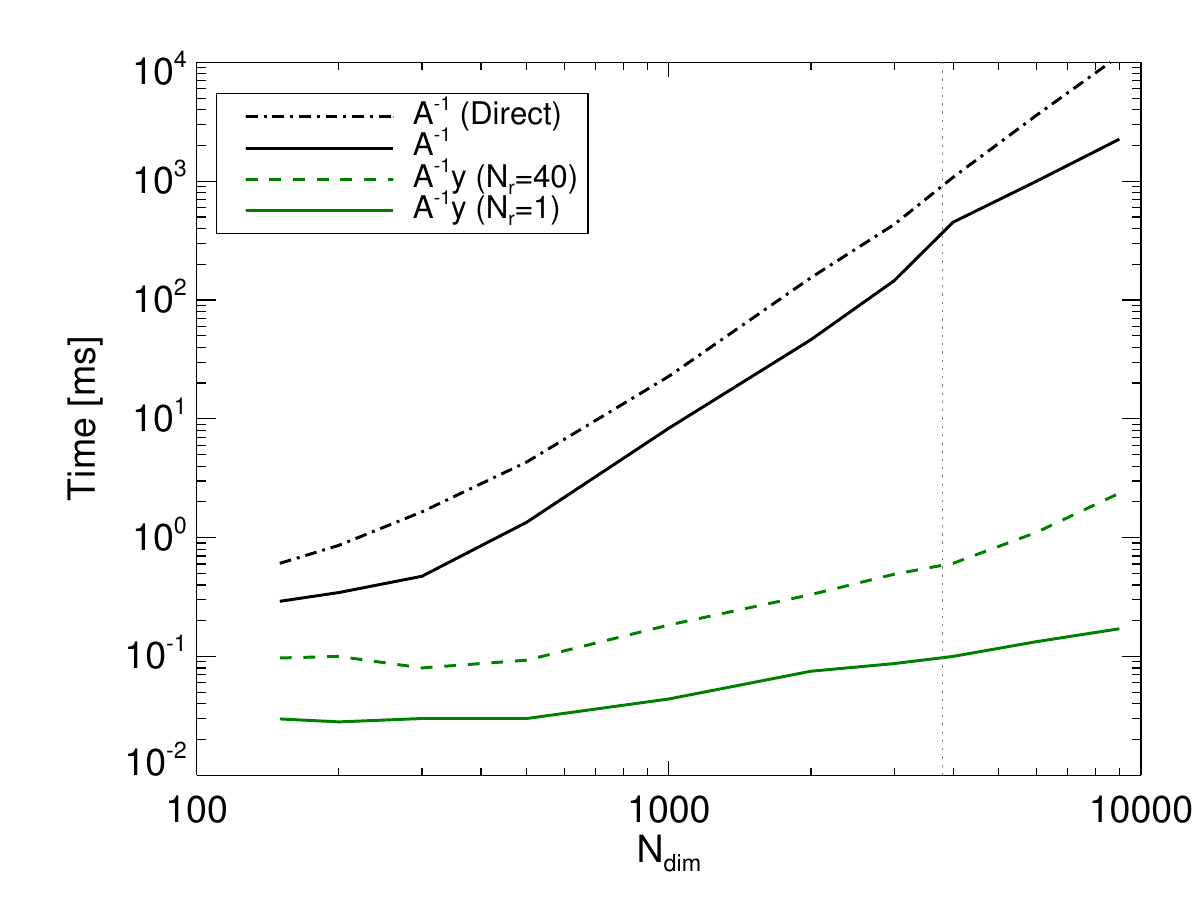}
\includegraphics[width=3.5in]{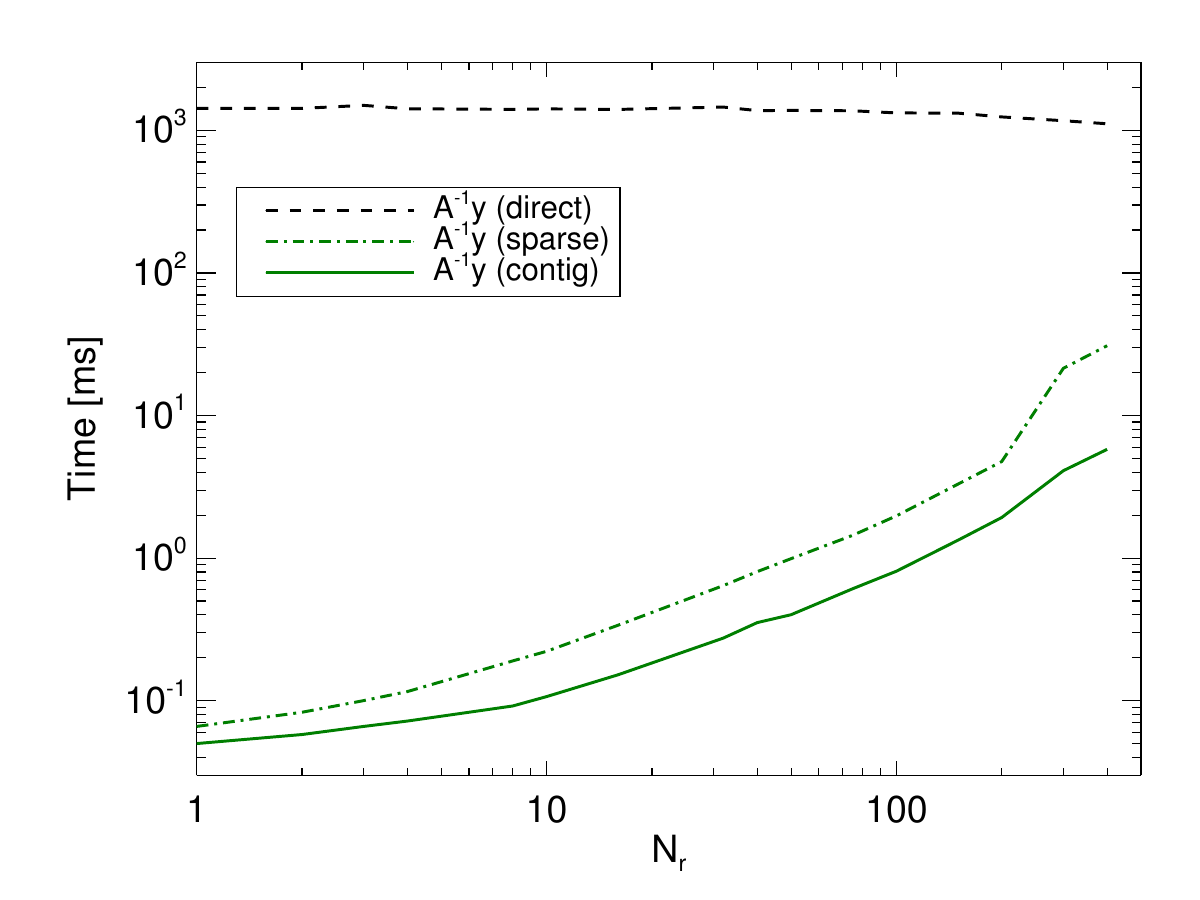}
\end{center}
\caption{Time required for inversion of submatrix $A$ of $M$, where $M$ is $N_{dim}\times N_{dim}$.  $A$ has one row and column removed.  Direct inversion of $A$ via Cholesky factorization is several times slower than the expression in Eq. (\ref{eq:Ainv}).  Computation of $A^{-1}y$ is faster still via Eq. (\ref{eq:Ainvy}).  The vertical line at $N_{dim}=3801$ shows the case relevant for LAMOST data.  All calculations are float64.}
\label{fig:speed_vs_ndim}
\end{figure}

The first step is to relate the inverse of a submatrix of $M$ to $M^{-1}$.  Suppose $A$, $B$, $B^T$, and $D$ are $p\times p$, $p\times q$, $q\times p$, and $q\times q$ matrices\footnote{row-major notation, so ``row''$\times$``column''.}, and $D$ is invertible. Let
\begin{equation} 
M=\left[{\begin{matrix}A&B\\
                       B^T&D\end{matrix}}\right] 
\end{equation}
so that $M$ is a $(p + q) \times (p + q)$ matrix.  It's inverse has blocks
\begin{equation} 
M^{-1}=\left[{\begin{matrix}P&Q\\
                           Q^T&U\end{matrix}}\right] 
\end{equation}
By the matrix inversion lemma,\footnote{This is the Schur complement of block $U$ of $M^{-1}$.} the inverse of $A$ is
\begin{equation} \label{eq:Ainv}
A^{-1} = P - QU^{-1}Q^T
\end{equation}
To verify, multiply both sides of Eq.~(\ref{eq:Ainv}) by $A$.  
\begin{equation} \label{eq:AinvA}
A^{-1}A = PA - QU^{-1}Q^TA
\end{equation}
Because $M^{-1}M = I$, 
\begin{eqnarray} 
Q^TA+UB^T & = & 0 \\
PA + QB^T & = & I
\end{eqnarray}
Combining these with Eq. (\ref{eq:AinvA}) verifies the identity, as long as $U$ is invertible.  While Eq. (\ref{eq:Ainv}) is faster to evaluate than a direct inverse (See Figure \ref{fig:speed_vs_ndim}) it is usually advantageous to set up the problem so that $A^{-1}$ is never evaluated.  

\subsection{Inverse of a submatrix times a vector}
In this section we consider the product of $A^{-1}$ and a vector $y$, and mention a few tricks to expedite the calculation.  Building on Eq. (\ref{eq:Ainv}) we have
\begin{equation} \label{eq:Ainvy}
A^{-1}y = Py - QU^{-1}Q^Ty
\end{equation}
Let's also assume we are going to apply this equation many times, so it is practical to pre-compute $M^{-1}$ and $M^{-1}y$.  Let $Y$ be the full vector of length $p+q$, and $y = Y[k]$, where $k$ is the index list of pixels to keep, and $y_r = Y[r]$, with $r$ the list of pixels to remove.  Note that 
\begin{equation}
Q^Ty = (M^{-1}Y)[r] - Uy_r
\end{equation}
The right hand side evaluates much faster than $Q^Ty$ for $N_r \ll N_k$. 

Next we need $U^{-1}Q^Ty$, but $U^{-1}$ is positive semi-definite, so we compute the product using Cholesky backsubstitution instead of directly inverting $U$.  $U^{-1}Q^Ty$ is a column vector of length $N_r$ and directly multiply $Q$ by it.  

\begin{figure}[t!]
\begin{center}
\includegraphics[width=3.5in]{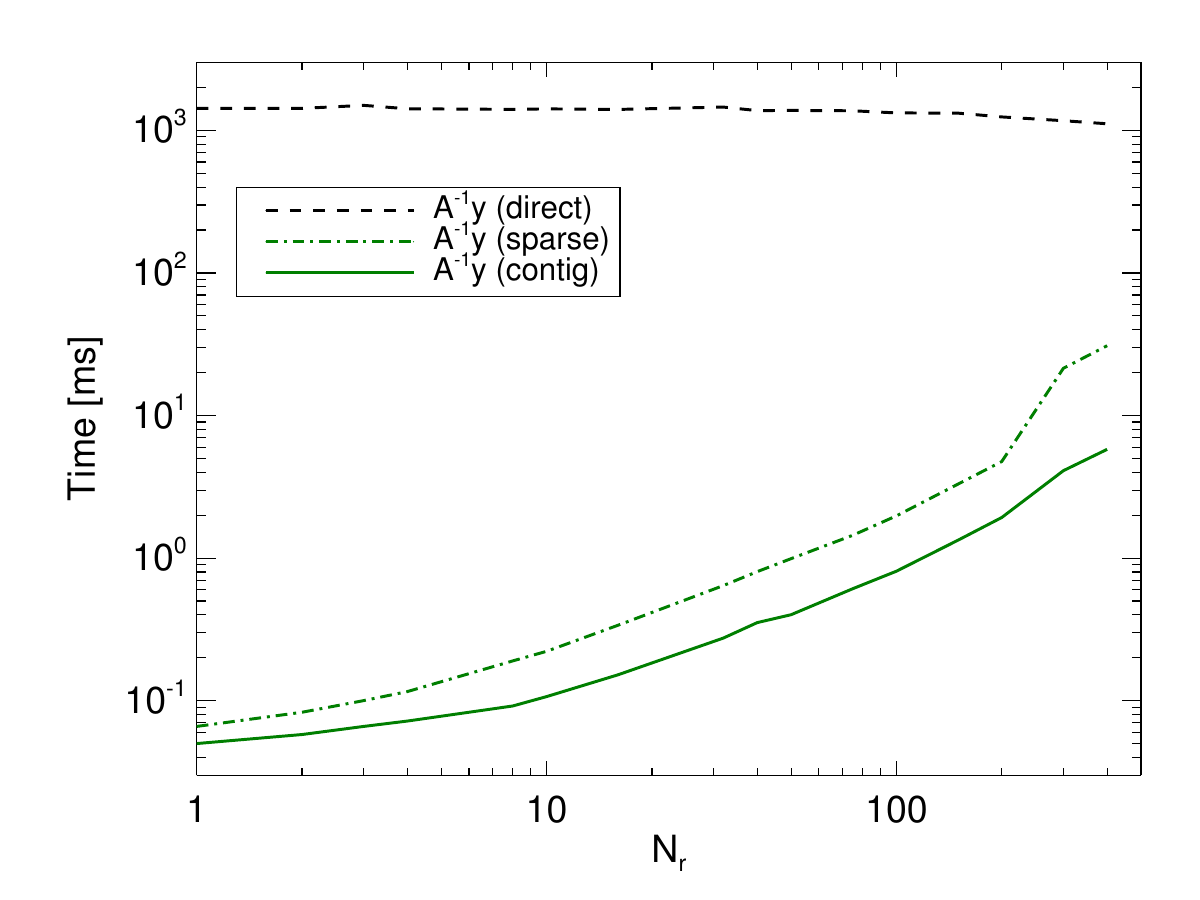}
\end{center}
\caption{Time required for computations involving a submatrix of a $4000\times4000$ matrix, $M$, and a column vector, $y$.  In all cases $A$ is a submatrix of $M$ with $N_r$ rows and columns removed.  Time to calculate $A^{-1}y$ with Cholesky factorization of $A$ to invert and then multiplying by $y$ is nearly constant as a function of $N_r$ (direct).  Using Equation \ref{eq:Ainvy} for $A^{-1}y$ is 3 orders of magnitude faster for $N_r=40$ in general (sparse) and even faster if the removed rows and columns are contiguous (contig).  Execution time is approximately $(8N_r + 40)\mu$s.  All calculations are float64. }
\label{fig:speed_vs_nr}
\end{figure}

These steps are fast enough that $Py$ would dominate the execution time, but $Py = (M^{-1}y)[k] - Qy[r]$. 
Using all of these shortcuts, the time required to evaluate $A^{-1}y$ for $N_k\sim 4000$ and $N_r\sim 40$ is 3 orders of magnitude faster than a direct inversion of $A$ (Figure \ref{fig:speed_vs_nr}).  

Most use cases of GSPICE involve either prediction of a single pixel for many spectra, or prediction of many pixels in a single spectrum.  This amounts to either $y^TA$ or $Ay$.  In other cases, many pixels for many spectra may be needed, but in this case matrix inversion generally doesn't dominate the time budget.  So for timing estimates at $N_\lambda=4000$, we can roughly say that inversions are 1ms and matrix multiplication consumes the rest of the time. 

%==================================================================
%=  FIGURE Speed   fig:speed_vs_ndim
%==================================================================

%==================================================================
%=  FIGURE Speed   fig:speed_vs_nr
%==================================================================

%

% Sherman, Jack; Morrison, Winifred J. (1949). "Adjustment of an Inverse Matrix Corresponding to Changes in the Elements of a Given Column or a Given Row of the Original Matrix (abstract)". Annals of Mathematical Statistics. 20: 621. doi:10.1214/aoms/1177729959

% \newpage
% \section{Software}
% \label{sec:software}
% We adopt the somewhat arbitrary convention that $10\sigma$ outliers are replaced, and their estimate is conditioned on pixels $< 5\sigma$ deviant.  In other words, pixels with $5<Z<10$ are not bad enough to discard, but not good enough to use in the estimate.  In Figure \ref{fig:repair} the repaired image is constructed in this way. 

% The GSPICE code delivers
% \begin{itemize}
% \item original spectrum
% \item original variance
% \item original mask
% \item GSPICE predicted mean
% \item GSPICE predicted variance
% \item repaired spectrum
% \item mask
% \end{itemize}

% \textbf{TODO: Fill in with examples! Also reference }

% In this section we provide pointers to software and examples. 

\end{document}

%% file: bangla_commands.tex
\def\bng{\bngx}

\font\bngx=bang10

\def\*#1*#2{o\null{#2}{#1}}

\def\d#1{\oalign{\smash{#1}\crcr\hidewidth{$\!$\rm.}\hidewidth}}

\def\sh#1{\setbox0=\hbox{#1}%
     \kern-.02em\copy0\kern-\wd0
     \kern.04em\copy0\kern-\wd0
     \kern-.02em\raise.0433em\box0 }